\documentclass[twocolumn,a4paper,useAMS,usenatbib]{mn2e}
\usepackage{natbib}

\usepackage{fmtcount}
\usepackage{threeparttable}
\usepackage{graphicx,subfigure}
\usepackage{multirow}
\usepackage{url}
\usepackage{amssymb}
\title[SN 2009ib in NGC 1559]{SN 2009ib: A Type II-P Supernova with an Unusually Long Plateau}
\author[K. Tak\' ats et al.]{K.~Tak\' ats$^{1,2}$\thanks{E-mail: ktakats@gmail.com}, G.~Pignata$^{1,2}$, M.~L.~Pumo$^{3,4,5}$, E.~Paillas$^1$, L.~Zampieri$^3$, \newauthor N.~Elias-Rosa$^3$, S.~Benetti$^3$, F.~Bufano$^{1,2}$, E.~Cappellaro$^3$, M.~Ergon$^6$, \newauthor M.~Fraser$^7$, M.~Hamuy$^{8,2}$, C.~Inserra$^9$, E.~Kankare$^9$, S.~J.~Smartt$^9$, \newauthor M.~D.~Stritzinger$^{10}$, S.~D.~Van Dyk$^{11}$,  J.~B.~Haislip$^{12}$, A.~P.~LaCluyze$^{12}$, \newauthor J.~P.~Moore$^{12}$ and D.~Reichart$^{12}$ \\
$^1$Departamento de Ciencias F\'isicas, Universidad Andr\'es Bello, Avda. Rep\'ublica 252, Santiago, Chile \\
$^2$Millennium Institute of Astrophysics, Santiago, Chile\\
$^3$INAF -- Osservatorio Astronomico di Padova, vicolo dell$'$Osservatorio 5, I-35122 Padova, Italy\\
$^4$Universit\`a di Catania, Dip. di Fisica e Astronomia, via S. Sofia 78, 95123 Catania, Italy \\
$^5$INAF -- Osservatorio Astronomico di Palermo, Piazza del Parlamento 1, 90134 Palermo, Italy \\
$^6$The Oskar Klein Centre, Department of Astronomy, AlbaNova, Stockholm University, SE-10691 Stockholm, Sweden\\
$^7$Institute of Astronomy, University of Cambridge, Madingley Road, Cambridge CB3 0HA, UK\\
$^8$Departamento de Astronom\' ia, Universidad de Chile, Casilla 36-D, Santiago, Chile\\
$^9$Astrophysics Research Centre, School of Mathematics and Physics, Queen's University, Belfast, Belfast BT7 1NN, UK \\
$^{10}$Department of Physics and Astronomy, Aarhus University, Ny Munkegade 120, DK-8000 Aarhus C, Denmark \\
$^{11}$IPAC/California Institute of Technology, 770 So. Wilson Ave., Pasadena, CA 91125 USA \\
$^{12}$University of North Carolina at Chapel Hill, Campus Box 3255, Chapel Hill, NC 27599-3255, USA\\
}

\begin{document}
\date{Accepted  Received ; in original form }

\pagerange{\pageref{firstpage}--\pageref{lastpage}} \pubyear{}

\maketitle

\label{firstpage}

\begin{abstract}

We present optical and near-infrared photometry and spectroscopy of SN~2009ib, a Type II-P supernova in NGC~1559. This object has moderate brightness, similar to those of the intermediate-luminosity SNe 2008in and 2009N. Its plateau phase is unusually long, lasting for about 130 days after explosion. The spectra are similar to those of the subluminous SN~2002gd, with moderate expansion velocities. We estimate the $^{56}{\rm Ni}$~mass produced as $0.046 \pm 0.015\,{\rm M}_{\sun}$. We determine the distance to SN~2009ib using both the expanding photosphere method (EPM) and the standard candle method. We also apply EPM to SN~1986L, a type II-P SN that exploded in the same galaxy. Combining the results of different methods, we conclude the distance to NGC~1559 as $D=19.8 \pm 3.0$~Mpc. We examine archival, pre-explosion images of the field taken with the Hubble Space Telescope, and find a faint source at the position of the SN, which has a yellow colour ($(V-I)_0 = 0.85$~mag). Assuming it is a single star, we estimate its initial mass as $M_{\rm ZAMS}=20\,{\rm M}_{\sun}$. We also examine the possibility, that instead of the yellow source the progenitor of SN~2009ib is a red supergiant star too faint to be detected. In this case we estimate the upper limit for the initial zero-age main sequence mass of the progenitor to be $\sim 14-17\,{\rm M}_{\sun}$. In addition, we infer the physical properties of the progenitor at the explosion via hydrodynamical modelling of the observables, and estimate the total energy as $\sim 0.55 \times 10^{51}$~erg,  the pre-explosion radius as $\sim 400\,{\rm R}_{\sun}$, and the ejected envelope mass as $\sim 15\,{\rm M}_{\sun}$, which implies that the mass of the progenitor before explosion was $\sim 16.5-17\,{\rm M}_{\sun}$.

\end{abstract}


\section{introduction}

Type II-P supernovae (SNe II-P) are by far the most common among the core-collapse SNe representing about $\sim 60$\% of them \citep[][volume-limited sample]{li}. They can be classified by the strong hydrogen lines in their spectra and by their characteristic light curves that show a near constant luminosity during the first $\sim 100$ days of evolution. 

SNe II-P with lower absolute magnitudes, slower expansion velocities and smaller $^{56}$Ni masses than the majority have been studied extensively. SN~1997D was the first subluminous SN analysed \citep{turatto_ba,benetti_1997D}, and was followed by e.g. 1999br \citep{hamuy_thesis, pastorello99br}, 2002gd \citep{spiro_2014}, 2003Z \citep{utrobin03Z,spiro_2014}, 2005cs \citep{takats2006,pastorello05csII}, and 2009md \citep{fraser09md}

Recently, a few analyses of SNe~II-P that have parameters falling in between those of the normal and subluminous objects have been published. SNe~2008in \citep{roy08in}, 2009js \citep{gandhi_09js}, and 2009N \citep{takats2014}  showed mixed properties, similarities to both groups. The observed properties of SNe~2008in and 2009N, such as their luminosities, spectra, expansion velocities were almost identical. By estimating the progenitor parameters at the explosion for SN~2009N, \citet{takats2014} found that the explosion energy and the estimated $^{56}$Ni mass of this object are between the typical values of the subluminous and normal SNe II-P. \citet{gandhi_09js} showed that the characteristics of SN~2009js place this object in between the subluminous SNe and SN~2008in. The studies of these intermediate SNe suggest that subluminous and normal SNe II-P do not form distinct groups.

Similar conclusions have been drawn by \citet{anderson_2014} and \citet{sanders_2014}, who  carried out studies that examined the light curves of a large number of SNe. \citet{anderson_2014} collected $V$-band light curves of 116 SNe, while \citet{sanders_2014} used  $grizy$-band photometry of Pan-STARRS1 objects. Both groups concluded that the light curve properties of SNe II, such as the absolute magnitude, plateau length, decline rate have a continuous distribution. 

SNe II-P are thought to emerge from stars with main sequence masses of $>8$~M$_{\sun}$. This is supported by cases, when red supergiant progenitors with masses between $8-16\,{\rm M}_{\sun}$ have been directly detected on archival, high-resolution pre-explosion images \citep[recent examples include ][]{smartt_rev,fraser09md,vandyk12aw,maund08bk}. Despite these studies, however, we still do not fully understand the causes of the observed diversity of these objects. The analysis of new and different objects can help to better understand their progenitor and the explosion mechanism.

In this paper we present data and analysis of an object, that seems to belong to this group of ``intermediate'' SNe~II-P, but also shows significant differences from those previously studied, further blurring the line between the subgroups of SNe II-P.
SN 2009ib was discovered by the Chilean Automatic Supernova Search (CHASE) project on August 6.30 UT 2009 in NGC~1559 \citep{2009ib_felfed}. This galaxy have hosted three other SNe, 1984J \citep[][type II]{1984j_discovery}, 1986L \citep[][type II-P]{1986l_discovery} and 2005df \citep[][type Ia]{2005df_discovery}. SN~2009ib was classified as a type II-P SN similar to SN 2005cs a week after its explosion on 2009 August 9.1 UT \citep{2009ib_classification}. 

This paper is organised as follows. In Sect.~\ref{sec_photometry} we present optical and near-infrared photometric observations, estimate the extinction towards the SN and examine its colour evolution. In Sect.~\ref{sec_spectroscopy} the optical and NIR spectra are shown and compared to those of other SNe II-P. We estimate the distance to SN~2009ib in Sect.~\ref{sec_distance} using the expanding photosphere method (EPM) and the standardized candle method (SCM), and apply EPM to SN~1986L, another SN~II-P that exploded in the same host galaxy. In this section we estimate the time of explosion of SN~2009ib as $t_0=55041.3$~MJD (2009 July 29.3~UT), the epoch we use throughout the paper. We calculate the bolometric luminosity and the produced $^{56}$Ni mass in Sect.~\ref{sec_bolometric}. In Sect.~\ref{sec_physical} we  examine the pre-explosion images of the SN site taken with the Hubble Space Telescope (HST) searching for the progenitor, and we also determine the progenitor properties via hydrodynamical modelling of the SN observables and compare the obtained physical parameters with those of other SNe~II-P. We compare the light curve and spectra of SN~2009ib to the non-LTE (non-local thermodynamic equilibrium) time-dependent radiative-transfer models of \citet{dessart2013}. Finally, in Sect.~\ref{sec_summary} the main results are summarized.


\section{photometry}\label{sec_photometry}

\begin{figure}
\includegraphics[width=84mm]{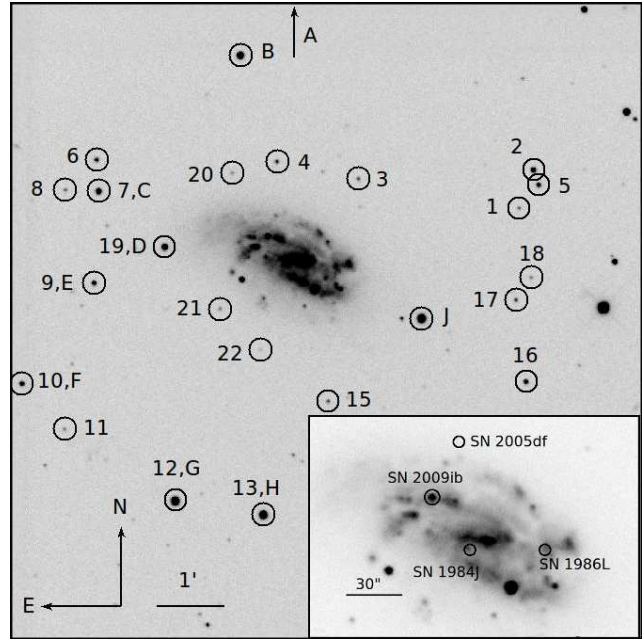}
\caption{The field of NGC~1559, the host galaxy of SN~2009ib. The stars used for photometric calibration are marked (see Tables \ref{seq_bvri} - \ref{seq_nir}). The image was taken on 2009 October 17 with the PROMPT5 telescope in V band. The insert shows the positions of all the SNe discovered in this galaxy, SN 1984J was a Type II, 1986L a Type II-P, 2005df a Type Ia SN.}
\label{snfield}
\end{figure}

Optical photometry was collected using multiple telescopes with $UBVRI$ and $u'g'r'i'z'$ filters, covering the phases between 13 and 262 days after explosion. The basic reduction steps of the images (such as bias-subtraction, overscan-correction, flat-fielding) were carried out using the standard {\sc iraf}\footnote{{\sc iraf} is distributed by the National Optical Astronomy Observatories, which are operated by the Association of Universities for Research in Astronomy, Inc., under the cooperative agreement with the National Science Foundation.} tasks. The photometric measurement of the SN was performed using the point-spread function (PSF) fitting technique via the {\sc snoopy} package\footnote{{\sc snoopy} is a package originally designed by F. Patat and later implemented in {\sc iraf} by E. Cappellaro.} in {\sc iraf}.
The calibration of the photometry was carried out by observing standard fields \citep{landolt,landolt2007,sloan_std_fields} on photometric nights. With the help of these images, magnitudes for a local sequence of stars (Fig.~\ref{snfield}, Tables~\ref{seq_bvri} and \ref{seq_ugriz}) on the SN field were determined and used to calibrate the SN measurements (Tables~\ref{lc_landolt} and \ref{lc_sloan}).

\begin{figure*}
\includegraphics[width=\textwidth]{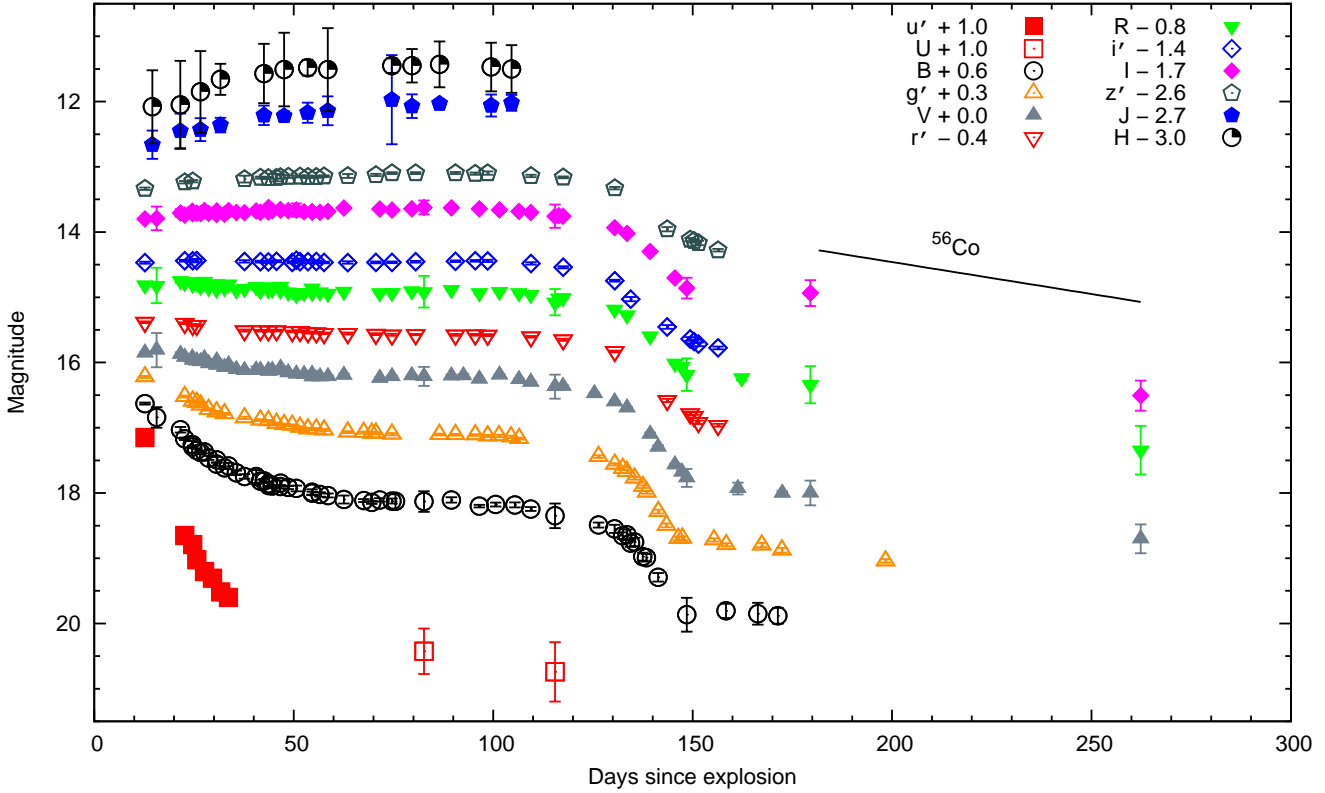}
\caption{Optical and near-infrared light curve of SN~2009ib. The phase is given relative to the estimated explosion epoch, $t_0=55041.3$~MJD (Sect.\ref{sec_explepoch}).}
\label{lc}
\end{figure*}

Near-infrared photometry was obtained using the Rapid Eye Mount telescope (REM) in $JH$ bands. Dithered images of the SN field were taken in multiple sequences of five. The object images were dark- and flatfield-corrected, combined to create a sky images then the sky images were substracted from the object images. The images were then registered and combined. Photometry was carried out with PSF-fitting, also using the {\sc snoopy} package. In order to calibrate the measurements, magnitudes of local sequence of stars were obtained from the 2MASS catalogue \citep[Table \ref{seq_nir};][]{2mass_catalogue}. The $JH$ magnitudes of the SN can be found in Table~\ref{lc_nir} and in Fig.~\ref{lc}.

SN~2009ib was discovered $\sim 8$~days after explosion ($\pm 2$~days, Section~\ref{sec_explepoch}), at the beginning of its plateau phase. On the plateau the magnitudes in $Vr'Ri'Iz'$ bands are roughly constant. In $u'Bg'$ bands the magnitudes decrease more rapidly in the first $\sim 40$~days, while in the $JH$ bands the brightness increases for $\sim 75$~days (Fig.~\ref{lc}).

The plateau phase of this SN is unusually long, longer than that of most SNe II-P. After the plateau the brightness drops almost $2$ magnitudes in $\sim 30$ days. Unfortunately, the tail phase is less well-sampled, with the last measurement at $262$~days. This phase is powered by the radioactive decay of $^{56}$Co to $^{56}$Fe. The expected decline rate, assuming full trapping of the gamma-rays, is $0.98$~mag/100d \citep{patat_1994}. Between days $160$ and $262$ we measure the decline rate in the bolometric light curve (Sect.\ref{sec_bolometric}) as $1.12 \pm 0.05$~mag/100d, while in $V$ band as $0.78 \pm 0.05$ mag/100d, in $R$ band as $1.14 \pm 0.09$ mag/100d.
Note, that number of data points to fit the light curve tail is low in every band and our observations only cover the beginning of the tail phase, therefore these measurements are quite uncertain.

\begin{figure}
\includegraphics[width=84mm]{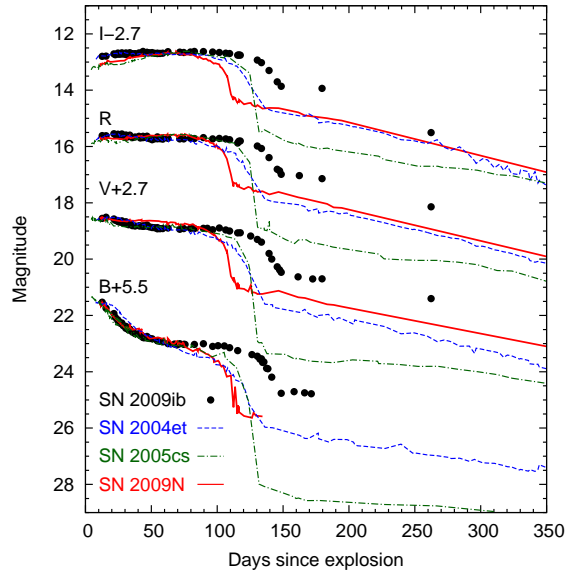}
\caption{Comparison of the $BVRI$ light curves of SNe II-P. The light curves were vertically shifted so their peak magnitudes match those of SN~2009ib in each band. The explosion epochs of the comparison SNe are very well-constrained (Table~\ref{otherSN}). The diversity of the shapes is reflected mostly in the length of the plateau and the drop to the tail. SN~2009N (red) has the same brightness as SN~2009ib (black circles), however its plateau is shorter. SNe~2005cs (green) and 2004et (blue) have about the same plateau length, but their absolute brightness differ by more than 2 magnitudes and the drop in the end of the plateau of SN~2005cs is greater.}
\label{lc_comp}
\end{figure}

Fig.~\ref{lc_comp} compares the shape of the light curves of a sample SNe II-P. We included in this sample the subluminous SN~2005cs, the normal SN~2004et, and the intermediate SN~2009N, which had a brightness similar to that of SN~2009ib. Some important parameters of the SNe used for comparison throughout the paper can be found in Table \ref{otherSN}. Fig.~\ref{lc_comp} shows that the main differences in the light curves are the length of the plateau and the brightness of the tail relative to the plateau. SN~2009ib has a very long plateau and a relatively small drop to the tail. SN~2009N has a significantly shorter plateau, while the drop to the tail is about the same as for SN~2004et. The normal SN~2004et and the subluminous SN~2005cs has similar plateau lengths, while the drop to the tail is greater for SN~2005cs.  

In order to compare the plateau length of SN~2009ib to that of other SNe, we determined the epoch of the middle of the transition from the plateau to the tail ($t_{PT}$) via fitting the analytic function of \citet{olivares_scm} to the $V$ band light curves of the SNe. Note, that \citet{anderson_2014} used the same method on a much larger sample to estimate $t_{PT}$. The difference between $t_{PT}$ and the explosion epoch ($t_0$) of the comparison SNe can be found in Table~\ref{otherSN}. In the case of SN~2009ib we obtained $t_{PT}-t_0=141 \pm 2$~days, which is about $8-33$ days longer than those of the comparison SNe.

\subsection{Reddening and Colour Curves}\label{sec_reddening}

The galactic extinction in the direction of SN~2009ib is low, $E(B-V)_{\rm MW}=0.0257 \pm 0.0002$~mag \citep{reddening_recalib}.
Using the classification spectrum taken with the Gemini South Telescope (see Sect. \ref{sec_spectroscopy}), we measured the equivalent width of the  Na~{\sc i}~D line at the redshift of the host galaxy as $EW({\rm NaD})=0.838 \pm 0.018$~\AA~(the components of the multiplet are not resolved in the spectrum). At the redshift of the Milky Way Na~{\sc i}~D absorption cannot be detected in the spectrum, in agreement with a low $E(B-V)_{\rm MW}$.  
Using the equation of \citet{poznanski2012}, the equivalent width of Na\,{\sc i}~D leads to the host galaxy reddening of $E(B-V)_{\rm host}=0.131 \pm 0.025$~mag. 
Together with the Milky Way component we estimate the total reddening as $E(B-V)_{\rm tot}=0.16 \pm 0.03$~mag. Note, however, that in several cases the Na~{\sc i}~D feature was shown to be an unreliable indicatior of extinction \citep[see e.g.][]{poznanski2011, phillips_extinction}.

\begin{figure}
\includegraphics[width=84mm]{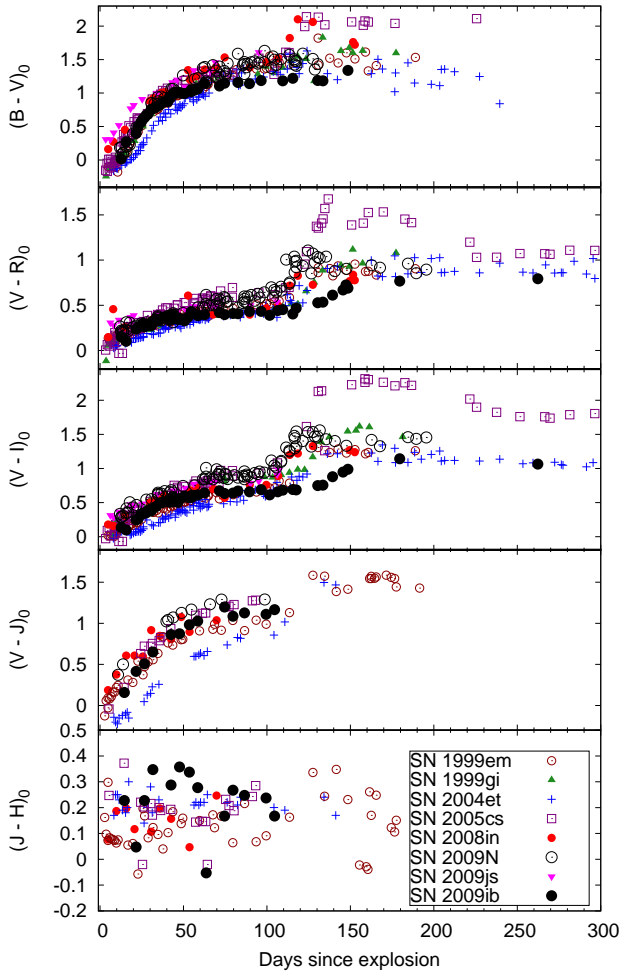}
\caption{The reddening corrected colour curves of SN~2009ib in comparison with those of other SNe II-P.}
\label{colour}
\end{figure}

Fig. \ref{colour} shows the $(B-V)_0$, $(V-R)_0$, $(V-I)_0$, $(V-J)_0$, and $(V-H)_0$ colours of SN~2009ib, corrected with the reddening above, and compared  with the colour curves of other SNe II-P.
The colours of SN~2009ib do not differ significantly from those of other II-P SNe.

During the second half of the plateau phase the colours of the SNe in our sample remain roughly constant, but when this phase ends they become significantly redder. The colour of SN~2009ib is mostly constant about $10-30$ days longer than those of the other SNe and only then becomes as redder, which is a manifestation of the longer plateau phase. During the nebular phase the colours are similar to those of SN~2004et.

\begin{table}
 \centering
 \begin{minipage}{84mm}
  \caption{Explosion epoch, reddening, distance and plateau length of the SNe used for comparison throughout the paper.}
  \label{otherSN}
  \begin{tabular}{@{}lccccc@{}}
  \hline
  \hline
SN & $t_0$  & $E(B-V)$ & $D$ & $t_0-t_{PT}^a$  & Ref. \\
& (MJD) & (mag) & (Mpc) & (days) & \\
\hline
1999em & 51476.5 (1.0) & 0.10 (0.05) & 11.7 (1.0) & 112 (3) & 1,2,3 \\
1999gi & 51517.8 (3.0) & 0.21 (0.10) & 11.1 (2.0) & 128 (5) & 4 \\
2002gd & 52551.5 (2.0) & 0.06 (0.01) & 37.5 (3.5) & 113 (7) & 5,6 \\
2003Z &  52664.5 (4.5) & 0.04 (0.01) & 21.9 (1.6) & 133 (8)  & 5 \\ 
2004et & 53270.0 (0.2) & 0.41 (0.01) & 4.8 (0.4)  & 123 (4) & 7,8 \\
2005cs & 53548.5 (0.5)  & 0.05 (0.01) & 8.4 (0.7) & 125 (6) & 9,10 \\
2008in & 54825.1 (0.8) & 0.10 (0.10) & 12.2 (1.9) & 108 (3) & 11  \\
2009N  & 54847.6 (1.2) & 0.13 (0.02) & 21.6 (1.1) & 109 (2) & 12 \\
2009js & 55115.4 (5.5) & 0.36 (0.12) & 21.7 (1.8) & 117 (7) & 13 \\ 
\hline				  
\end{tabular}			  
 \begin{tablenotes}
       \item[a]{References: (1) \citet{hamuy_epm}, (2) \citet{leonard_99em}, (3) \citet{leonard_99em_ceph}, (4) \citet{leonard_99gi}, (5) \citet{spiro_2014}, (6) \citet{poznanski_scm}, (7) \citet{takats2012},  (8) \citet{maguire_04et},  (9) \citet{pastorello05csII}, (10) \citet{vinko11dh}, (11) \citet{roy08in}, (12) \citet{takats2014}, (13) \citet{gandhi_09js}.}
       \item[b]{$^a$ $t_{PT}$ is the epoch of the middle of the transition from the plateau to the tail.}
           \end{tablenotes}
\end{minipage}
\end{table}


\section{spectroscopy}\label{sec_spectroscopy}

\begin{figure*}
\includegraphics[width=\textwidth]{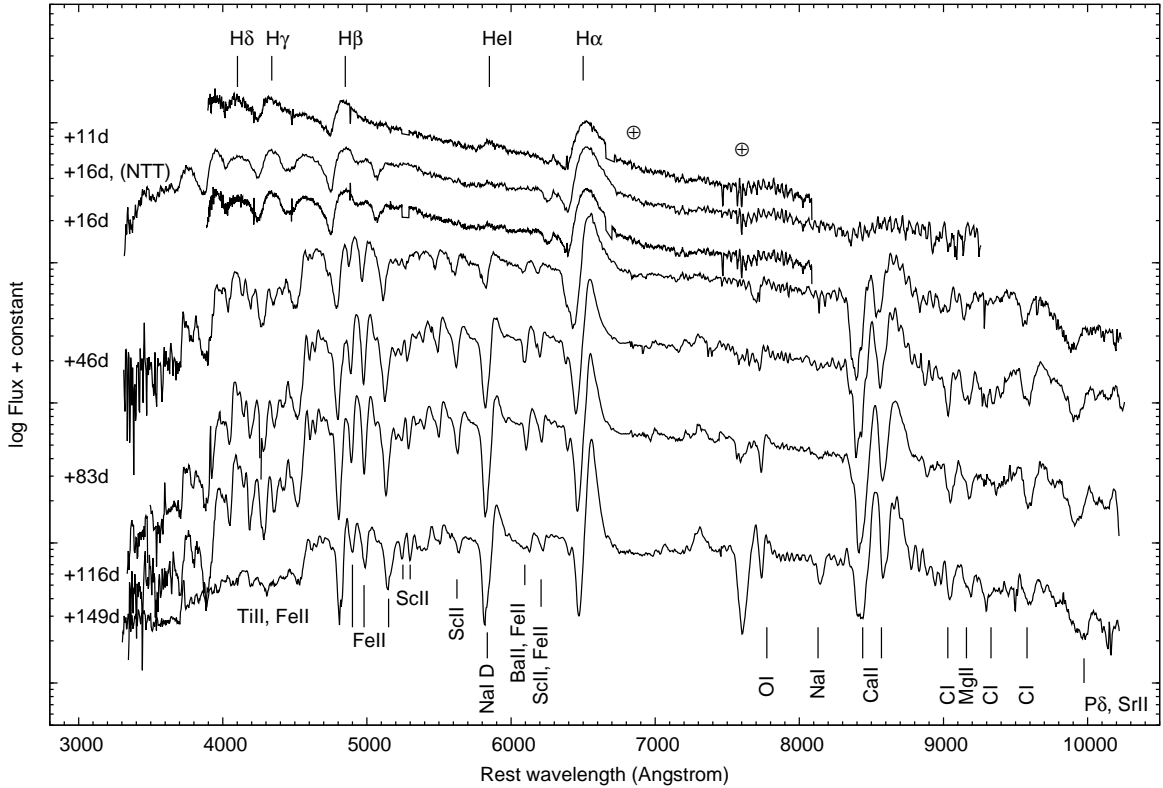}
\caption{The spectral evolution of SN~2009ib during its photosheric phase. Note, that in the case of the Gemini spectra the flux calibration is approximate (see text). In the figure we identify the strongest features present in the spectra. The wavelength positions of telluric features are marked with $\oplus$ symbol.}
\label{sp_evol}
\end{figure*}

\begin{figure*}
\includegraphics[width=\textwidth]{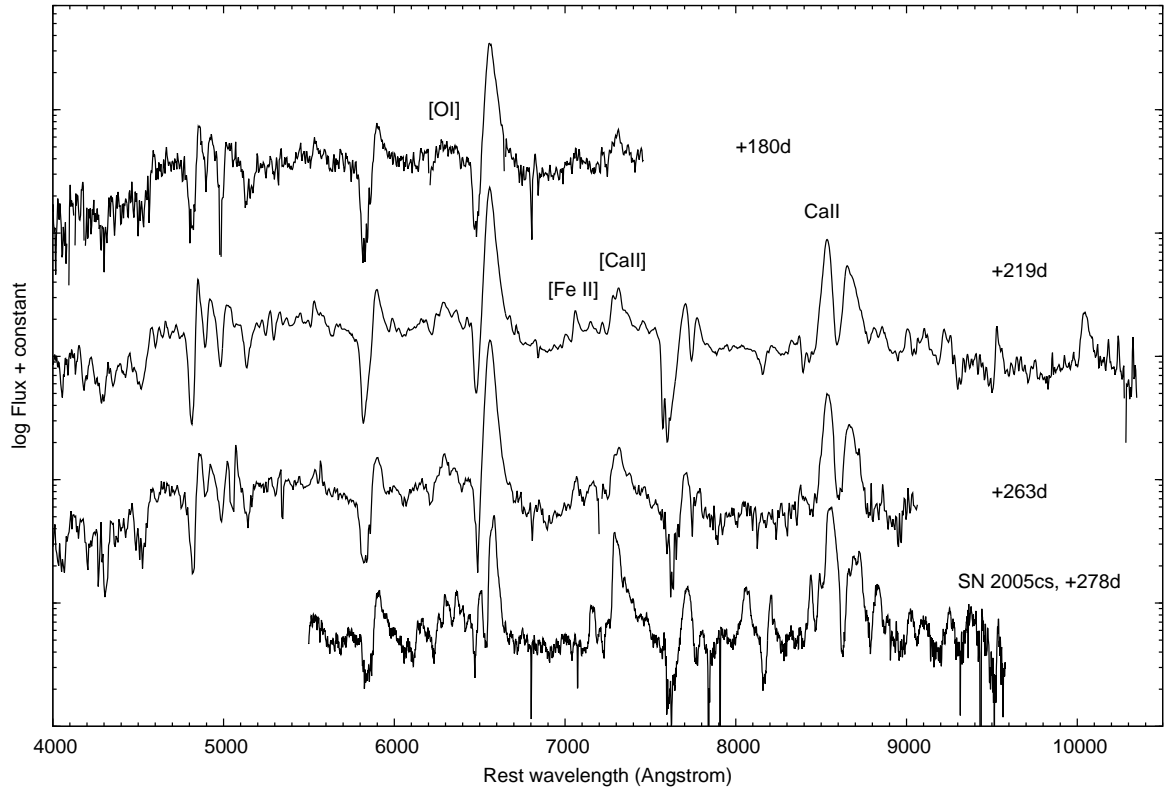}
\caption{The spectra of SN~2009ib during its tail phase. The lines that are typically present in the nebular phase spectra of SNe II-P are marked. A spectrum of the subluminous SN~2005cs at similar phase is also shown for comparison.}
\label{spec_nebu}
\end{figure*}

Optical spectroscopic observations of SN~2009ib were taken with the Gemini South Telescope (GST$+$GMOS), the New Technology Telescope (NTT$+$EFOSC2)  and the Very Large Telescope (VLT$+$FORS2) between days $11$ and $263$ after explosion (for the summary of the observations see Table \ref{logofsp}). The image reduction was carried out using standard {\sc iraf} tasks. After the basic reduction steps the spectrum was extracted and wavelength calibrated using comparison lamp spectra. We used spectra of standard stars observed on the same night in order to flux calibrate the SN spectrum. The flux was then checked against the photometric measurements taken at a close epoch and corrected using a scaling factor if necessary. The spectra are shown in Fig.~\ref{sp_evol} and \ref{spec_nebu}. In the case of the first  spectrum taken with GST  there was no flux standard star observed, therefore we used the standard taken during the night of our second GST observation, on August 14, for flux calibration. The continuum of both spectra, indicate unreasonably high temperatures, in disagreement with the photometric measurements. Since we have another spectrum taken by NTT at the same epoch as the second GST spectrum, we adjusted the continuum of both of the GST spectra with the help of this NTT spectrum, resulting in more realistic but quite uncertain continua (Fig.~\ref{sp_evol}).

The classification spectrum of SN~2009ib was taken on Aug. 9.1 UT with the GST \citep{2009ib_felfed}. Lines of the Balmer-series of H\,{\sc i} and He~{\sc i} $\lambda5876$ are present. According to the SNID program \citep[Supernova Identification,][]{SNID} the spectrum is similar to that of SN 2005cs about a week after explosion. By day 16  metallic lines had appeared such as Fe\,{\sc ii}, Ti\,{\sc ii}, Si\,{\sc ii}, Ca\,{\sc ii}, while He\,{\sc i} was not visible any more. The third spectrum was taken at the phase of $46$ days, when the metallic lines were much stronger and lines of Na\,{\sc i}~D, Sc\,{\sc ii}, Ba\,{\sc ii} and O\,{\sc i}  also appeared. During the rest of the plateau phase these lines persisted, and also remained present after the luminosity drop, at the phase of $263$ days. In our last two spectra, taken on days $219$ and $263$, nebular lines are also detected. The emission component of the Ca\,{\sc ii} IR triplet became prominent, and forbidden lines, such as [Ca\,{\sc ii}] $\lambda\lambda$7291,7323, [O\,{\sc i}] $\lambda\lambda$6300,6364, [Fe\,{\sc ii}] $\lambda\lambda$7155,7172 are present (Fig.\ref{spec_nebu}).

\begin{figure}
\includegraphics[width=84mm]{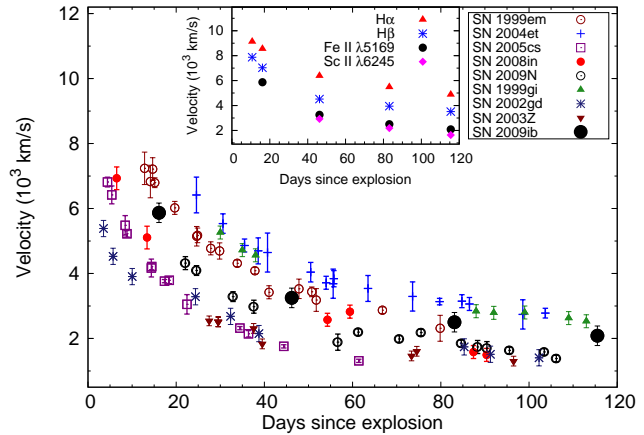}
\caption{Comparison of the expansion velocities of SN~2009ib measured from the Fe\,{\sc ii} $\lambda$5169 to those of the normal SNe II-P 1999em, 1999gi and 2004et, the subluminous SNe 2002gd, 2005cs, 2003Z and the intermediate luminosity SNe 2008in and 2009N. The insert shows Doppler-velocities from the absorption minima of H$\alpha$, H$\beta$, Fe\,{\sc ii} $\lambda$5169 and Sc\,{\sc ii} $\lambda$6245 lines of SN~2009ib.}
\label{line_vel}
\end{figure}

The spectra of SN~2009ib have narrow lines with low velocities. We measured the Doppler-velocities of H$\alpha$, H$\beta$, Fe\,{\sc ii} $\lambda$5169 and Sc\,{\sc ii} $\lambda$6245 lines during the photospheric phase by fitting a gaussian to their absorption minima. The velocity curves are shown in the insert of Fig. \ref{line_vel}. We compare the velocities of SN~2009ib to those of other SNe in Fig.\ref{line_vel} measured from the absorption minimum of the Fe\,{\sc ii} $\lambda$5169 line. The velocity curves of these objects have a high scatter, the ``normal'' SNe II-P, like SNe 1999em, 1999gi and 2004et have higher velocities than the rest of the sample. The velocities of SN~2009ib are somewhat higher than those of the intermediate luminosity SNe~2008in and 2009N.

\begin{figure*}
\includegraphics[width=\textwidth]{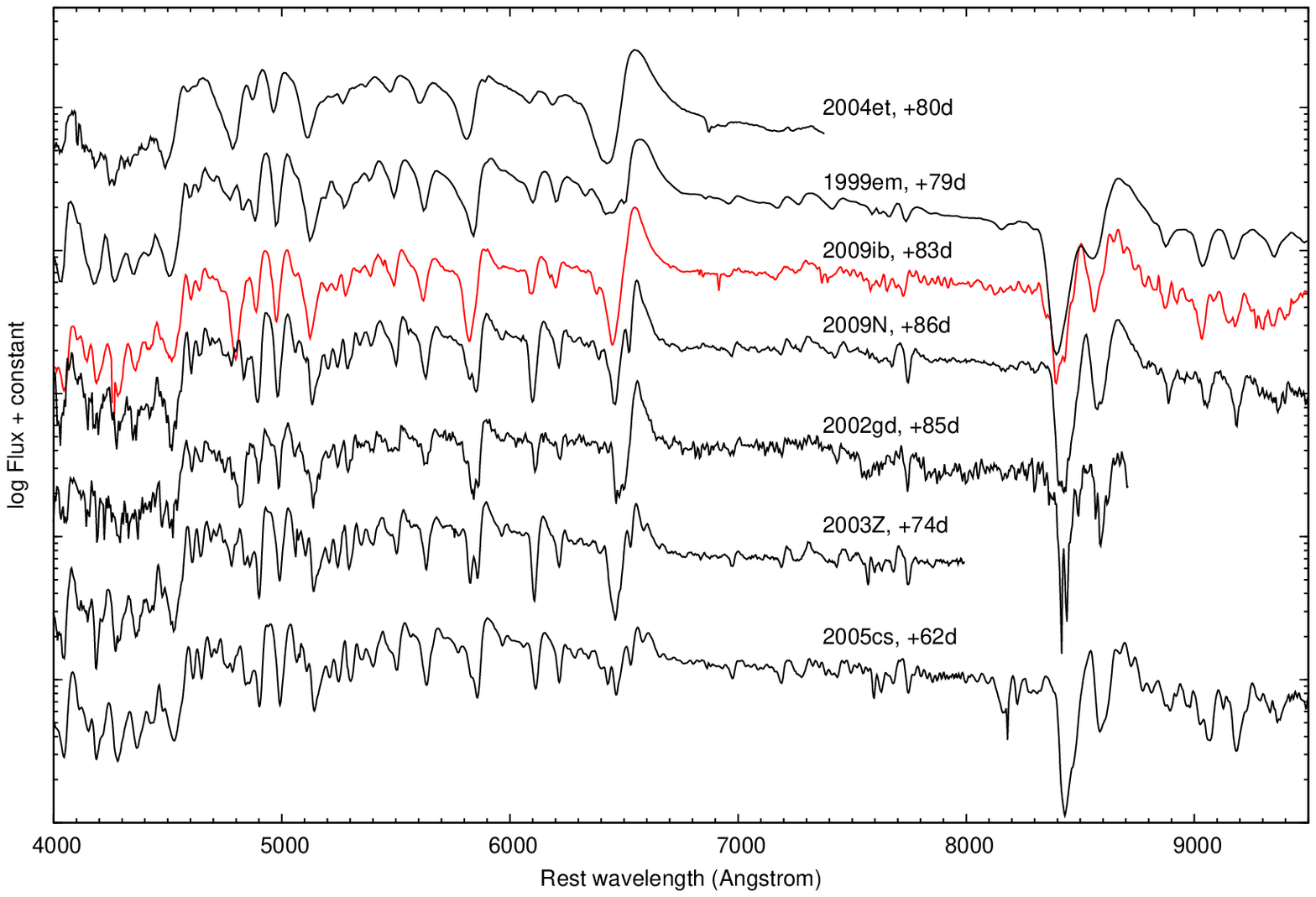}
\caption{Comparison of the spectra of selected SNe II-P at similar phases. The spectra are sorted by the brightness of the SNe, the brightest one being on the top, and includes those of the normal SNe 1999em, 2004et, the subluminous SNe 2002gd, 2003Z, 2005cs and the intermediate luminosity SN 2009N.  }
\label{spec_osszehas}
\end{figure*}

In Fig. \ref{spec_osszehas} we compare the $+83$ day spectrum of SN~2009ib with the spectra of normal and subluminous SNe taken at similar phases. The shape, width and relative strength of the features in the spectrum of SN 2009ib, especially the lines of H\,{\sc i}, Ba\,{\sc ii} and Fe\,{\sc ii}, appear to be the most similar to those of the subluminous SN~2002gd and differ somewhat from those of SN~2009N, and from those of the other subluminous SN included in this comparison. 
It demonstrates that the observational properties of SNe II-P are diverse, they do not form clearly distinguishable groups.

 \begin{table*}
 \centering
 \begin{minipage}{\textwidth}
  \caption{Summary of the spectroscopic observations.}
  \label{logofsp}
  \begin{tabular}{@{}cccccc@{}}
  \hline
  \hline
Date & MJD & Phase$^a$ & Instrument set-up & Wavelength range  \\
     &  & (days) & & (\AA) \\
  \hline
09/08/2009 & 55052.4 & 11.1 & GST + GMOS + R400 & 3900-8100 \\
13/08/2009 & 55057.3 & 16.0 & NTT + EFOSC2 + gr\#11 + gr\#16 & 3500-10100  \\
13/08/2014 & 55057.4 & 16.1 & GST + GMOS + R400 & 3900-8100 \\
12/09/2009 & 55087.4 & 46.1 & NTT + EFOSC2 + gr\#11 + gr\#16 & 3500-10100  \\
14/09/2009 & 55089.3 & 48.0 & NTT + SOFI + GB + GR & 9500-25200 \\
19/10/2009 & 55124.3 & 83.0 & NTT + EFOSC2 + gr\#11+gr\#16 & 3500-10100\\
20/10/2009 & 55125.3 & 84.0 & NTT + SOFI + GB & 9500-16400 \\
22/11/2009 & 55157.2 & 115.9 & NTT + EFOSC2 + gr\#11 + gr\#16 & 3500-10100  \\
24/12/2009 & 55190.2 & 148.9 & NTT + EFOSC2 + gr\#11 + gr\#16 & 3500-10100  \\
24/01/2010 & 55221.2 & 179.9 & NTT + EFOSC2 + gr\#11 & 3500-9500 \\
04/03/2010 & 55260.1 & 218.8 & VLT + FORS2 + 300V + 300I & 3500-11000 \\
17/04/2010 & 55304.0 & 262.7 & NTT + EFOSC2 + gr\#13 & 3500-9500 \\
\hline				  
\end{tabular}			  
 \begin{tablenotes}
       \item[a]{$^a$ relative to the estimated date of explosion, $t_0=55041.3$~MJD}
     \end{tablenotes}
\end{minipage}
\end{table*}

\subsection{NIR spectra}

Two NIR spectra were taken with NTT (+SOFI), on days $+48$ and $+84$ (Table \ref{logofsp}). The reduction steps were carried out using standard {\sc iraf} packages. Each night several pairs of spectra were taken at different positions along the slit. These pairs were subtracted from each other in order to remove the sky background, the subtracted images were then added together. The SN spectrum was extracted from the co-added image. The wavelength calibration was carried out using arc lamp spectra that was taken before and after the SN spectra. The strong telluric features were removed with the help of the spectrum of a G-type telluric standard star observed close in time, with the same instrumental setup and at similar airmass as the SN. 
We also used the spectrum of the telluric standard for the flux calibration of the SN spectra that were later checked against the NIR photometry from the nearest epoch and corrected when necessary. 

\begin{figure*}
\includegraphics[width=\textwidth]{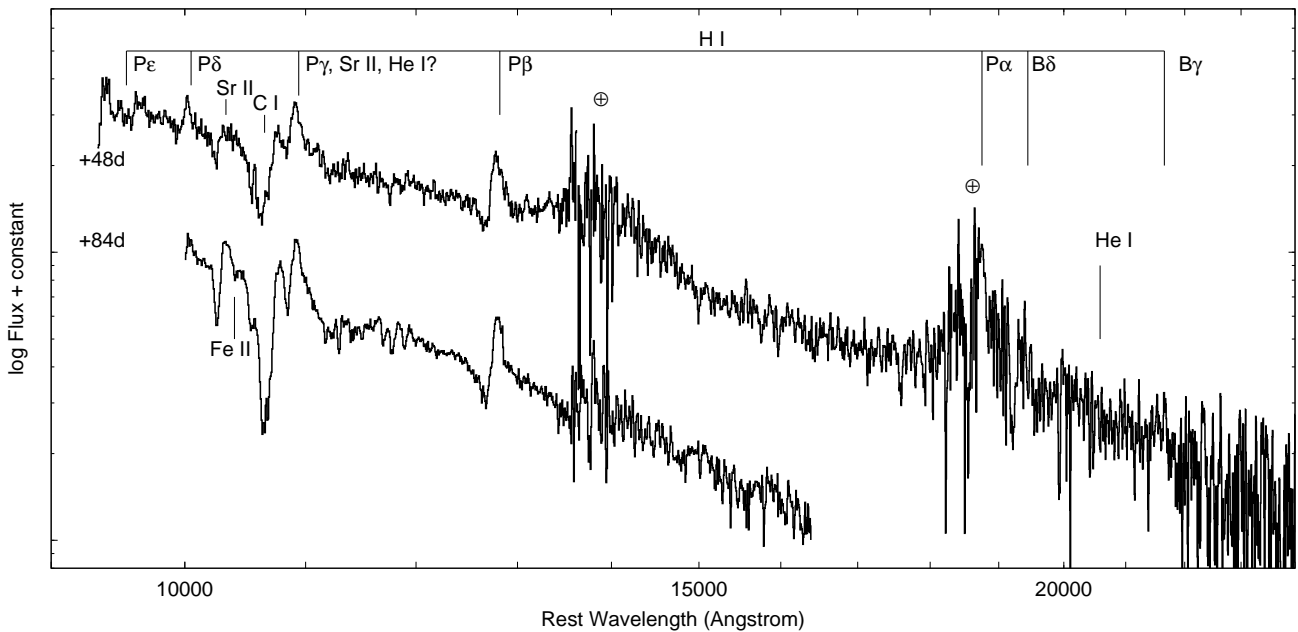}
\caption{The near-infrared spectra of SN~2009ib, taken $48$ and $84$ days after explosion.}
\label{nir_sp}
\end{figure*}

\begin{figure}
\includegraphics[width=84mm]{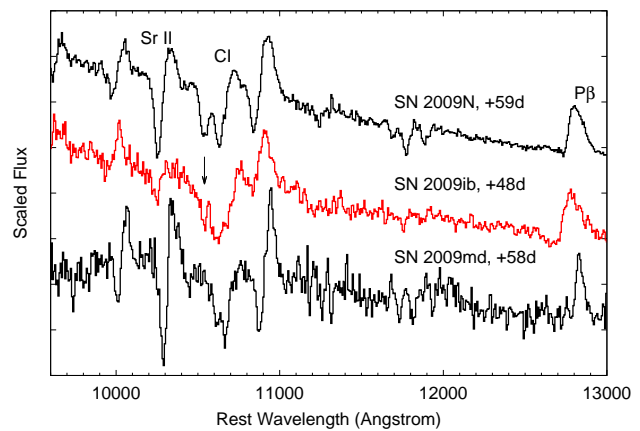}
\caption{Comparison of the $+$48d NIR spectrum of SN~2009ib (red) to those of SNe~2009N and 2009md at similar phases. The position of an interesting feature at $10550$~\AA, that was identified by \citet{takats2014} as either a high-velocity component of He\,{\sc i}~$\lambda$10830 or a Si\,{\sc i} feature is marked with an arrow.}
\label{nir_comp}
\end{figure}

The NIR spectra of SN~2009ib are shown in Fig. \ref{nir_sp}. We marked the position of the spectral lines commonly seen in NIR spectra of SNe II-P. Along with the H\,{\sc i} features, lines of Sr\,{\sc ii}, C\,{\sc i} and Fe\,{\sc ii} are visible. The first NIR spectrum covers also the range redwards from 16000~\AA, where we marked the position of He\,{\sc i} $\lambda20580$, however the line is not visible. In Fig. \ref{nir_comp} we compare the NIR spectrum of SN~2009ib taken at phase $+84$ days to those of SN~2009N and the subluminous SN~2009md \citep{fraser09md} at similar phases. While the Sr\,{\sc ii} lines seem to be weaker, the C\,{\sc i}~$\lambda 10691$ feature in the spectrum of SN~2009ib is significantly stronger and broader than those of the other two. The shapes of the emission components of the H\,{\sc i} lines of SNe~2009N and 2009ib are similar, but somewhat wider than those of SN~2009md. The feature at $10550$~\AA~\citep[which was present in the spectrum of SNe 2009N and 2008in, ][]{takats2014} also appears in the spectra of SN~2009ib (marked with an arrow in Fig.\ref{nir_comp}), however less pronounced, maybe due to the strong, wide C\,{\sc i} feature next to it. This line was identified by \citet{takats2014} as either a high-velocity component of He\,{\sc i} $\lambda10830$ or a Si\,{\sc i} feature, however they were unable to draw a definite conclusion (see their Fig.16).

\section{Distance}\label{sec_distance}

The distance to the host galaxy of SN~2009ib, NGC~1559, has been measured several times. The results, however, show a high scatter,  falling in the range between $12.6$ and $22.0$~Mpc (see Sect.~\ref{sec_ave}). 

There are two methods that allow us to measure distances of SNe~II-P directly, EPM \citep{epm_ref} and SCM \citep{hamuy_scm}. We apply both methods on the data of SN~2009ib (Sect.\ref{sec_epm} and \ref{sec_scm}) and EPM on the data of SN~1986L, an earlier SN that exploded in the same galaxy (Sect.\ref{sec_86L}).

\subsection{EPM}\label{sec_epm}

EPM is a variant of the Baade-Wesselink method, its application requires photometric and spectroscopic monitoring throughout the first half of the plateau phase. A great advantange of this method is that it does not require calibration using objects with known distances. Assuming that the optically thick ejecta expands homologously and radiates as a diluted blackbody, the method relates the apparent radius of the photosphere to its physical radius. The method has been discussed extensively in the literature, so for details we refer the reader to the relevant papers \citep[e.g. ][]{hamuy_thesis, leonard_99em, D05}.

To calculate the physical radius of the photosphere we used equation (1) of \citet{hamuy_epm}, and applied it to the $BVI$ band photometry. We adopt the velocities measured from the absorption minimum of the Fe\,{\sc ii}~$\lambda5169$ feature  as the photospheric velocities. We interpolated them to the epochs of the photometry using the relation given in the equation (2)  of \citet{takats2012}. The dilution factors can be determined as a function of the photospheric temperature from atmospheric model calculations. So far there have been two such studies, those of \citet{E96} and \citet{D05}. In this paper we adopted the more recent values, those of \citet{D05}. Note that the values of \citet{E96} are systematically lower, leading to shorter distances. The quantities calculated to apply EPM  can be found in Table \ref{epm}. The resulting distance is $D_{\rm EPM}=21.3 \pm 2.2$~Mpc, while the explosion date determined via EPM is $t_{0}=55041.3 \pm 3.1$~MJD.

\begin{table}
 \centering
 \begin{minipage}{84mm}
  \caption{The calculated parameters of SN~2009ib needed to apply EPM: the apparent radius ($\theta$), photospheric temperature ($T$), dilution factor ($\zeta(T)$), and the photospheric velocity ($v$), measured from the absorption minimum of Fe\,{\sc ii}~$\lambda5169$ and interpolated to the epochs of the photometry.}
  \label{epm}
  \begin{tabular}{@{}cccccc@{}}
  \hline
  \hline 
 MJD & ${t-t_0 \over 1+z}^a$ & $\theta$ & $T$ & $\zeta(T)$ & $v$ \\  
  & (days) & ($10^8$ ${{\mathrm {km}} \over {\mathrm {Mpc}}}$) & (K) & & (km/s) \\ 
  \hline
55054.3 & 13.0 & 3.481 & 10820 & 0.52 & 6999 \\
55057.3 & 15.9 & 3.959 & 9278 & 0.54 & 6181 \\
55063.3 & 21.9 & 4.860 & 7578 & 0.62 & 5107 \\
55064.3 & 23.0 & 4.656 & 7622 & 0.61 & 4965 \\
55066.3 & 24.9 & 5.172 & 6780 & 0.68 & 4729 \\
55066.3 & 25.0 & 4.895 & 7097 & 0.65 & 4729 \\
55067.3 & 26.0 & 4.907 & 7110 & 0.65 & 4621 \\
55068.3 & 26.9 & 4.941 & 6956 & 0.66 & 4522 \\
55069.3 & 27.9 & 5.538 & 6199 & 0.75 & 4433 \\
55070.3 & 28.9 & 4.934 & 6867 & 0.67 & 4339 \\
55072.3 & 30.9 & 5.091 & 6497 & 0.71 & 4176 \\
55072.4 & 31.1 & 5.121 & 6579 & 0.70 & 4172 \\
55074.3 & 32.9 & 5.239 & 6205 & 0.75 & 4028 \\
55075.4 & 34.0 & 5.124 & 6540 & 0.71 & 3956 \\
55077.3 & 35.9 & 5.398 & 6022 & 0.77 & 3832 \\
55079.3 & 37.9 & 5.864 & 5512 & 0.87 & 3714 \\
\hline
\end{tabular}			  
 \begin{tablenotes}
       \item[a]{$^a$ relative to the estimated date of explosion, $t_0=55041.3$~MJD}
     \end{tablenotes}
\end{minipage}
\end{table}

\subsubsection{Confirming the date of explosion}\label{sec_explepoch}

In order to test the explosion epoch obtained via EPM, we estimated this date by matching the plateau phase spectra of SN~2009ib to those of other similar objects. We used the applications {\sc snid} and {\sc gelato}\footnote{\url{https://gelato.tng.iac.es}} \citep{gelato}, both of which compare the given spectrum to extended databases of SN spectra.

First, we selected a sample of comparison SNe that have good spectral covarage and their explosion epochs are well-constrained. Then we narrowed this sample by comparing their spectra to those of SN~2009ib, in order to filter out those objects that have significantly different spectral evolution. The final sample of templates contains  SNe 1999em, 2002gd, 2004et, and 2009N for {\sc snid}, and SNe 1999em, 2004et, 2012A and 2007od in the case of {\sc gelato}. We matched each plateau phase spectra of SN~2009ib to the templates of these comparison SNe one by one, and calculated the explosion epoch taking into account the quality of the fits, i.e. the {\it rlap} parameter in the case of {\sc snid} and the {\it Quality of Fit} parameter in the case of {\sc gelato}. The systematic errors are calculated as the RMS of the explosion epochs from the different template SNe.  In the case of {\sc snid} the mean value of the determined  explosion epochs was $t_{\rm snid}=55044.7 \pm 6.5\,{\rm (random)} \pm 1.3\,{\rm (systematic)}$. With {\sc gelato} the systematic error of the results is higher, we obtained  $t_{\rm gelato}=55040.3 \pm 7.7\,{\rm (random)} \pm 3.2\,{\rm (systematic)}$.
In spite of the high uncertainties due to the slow spectral evolution of SNe II-P during the plateau phase, these results confirm  the epoch obtained via EPM, which therefore we adopt as the explosion epoch of SN~2009ib.

\subsection{SCM}\label{sec_scm}

Another method to calculate distances to SNe II-P is SCM which is based on an empirical correlation between the brightness of the SN and its expansion velocity in the middle of the plateau phase \citep{hamuy_scm}. The method requires calibration using objects with well-known distances. The original version has been revised multiple times \citep{hamuy_scm2, nugent, poznanski_scm, olivares_scm}. Here we apply several versions, similarly as was done in \citet{takats2014}. 

\citet{poznanski_scm} calibrated the equation
\begin{eqnarray}
\label{poz_eq}
\mathcal{M_I} - \alpha \cdot \log \left( {v_{\rm Fe}(50\mathrm{d}) \over 5000} \right) +R_I((V-I) - (V-I)_0) - m_I\\
= -5 \cdot \log(H_0D)\nonumber
\end{eqnarray}
and defined $\mathcal{M_I}=-1.615 \pm 0.08$~mag, $\alpha=4.4 \pm 0.6$, $R_I=0.8 \pm 0.3$ and $(V-I)_0=0.53$~mag. In the case of SN~2009ib, we measured $m_I=15.368 \pm 0.020$~mag, $(V-I)=0.800 \pm 0.032$~mag, and estimated $v_{\rm Fe}(50\mathrm{d})=3247 \pm 200$~km$/$s. Assuming a Hubble constant $H_0=72~{\rm km~s^{-1}~Mpc^{-1}}$, we calculated the distance as $D_{\rm SCM,1}=21.48 \pm 2.00$~Mpc.

\citet{olivares_scm} examined the data of 37 nearby II-P SNe, and applied the same expression as \citet{hamuy_scm}:
\begin{equation}
m+\alpha \log(v_{\rm Fe}/5000)-\beta(V-I)=5\log H_0D + zp,
\end{equation}
but using the magnitudes and velocities measured 30 days before the middle of the transition phase ($t_{\rm PT}$).
They calibrated the formula using different bands, the values of $\beta$ and $zp$ can be found in Table 6 of \citet{olivares_scm}. In Section~\ref{sec_photometry} we determined $t_{\rm PT}=55182.7$~MJD ($141$~days after explosion). At  30 days before this date we measured the values $v(111{\rm d})=2010 \pm 300~{\rm km/s}$, $m_B=18.49 \pm 0.06$~mag, $m_V=17.10 \pm 0.03$~mag, $m_R=16.39 \pm 0.02$~mag and $m_I=15.99 \pm 0.02$~mag. These values led to the distances $23.70 \pm 3.50$~Mpc, $26.74 \pm 3.29$~Mpc and $28.09 \pm 2.93$~Mpc in $B$, $V$ and $I$ bands, respectively. The mean of these results is $D_{\rm SCM,2}=26.19 \pm 3.25$~Mpc.

We also apply the version of \citet{maguire_scm}, who extended the technique to the NIR range.
They used the same formula as \citet{poznanski_scm} (equation~\ref{poz_eq}). Their calibration in $J$ band yielded the values of $\mathcal{M_J}=-2.532 \pm 0.250$~mag and $\alpha=6.33 \pm 1.20$. \citet{maguire_scm} chose to calculate with the value of $R_V=1.5$, obtained by \citet{poznanski_scm}, instead of using it as a fitting parameter. In the case of SN~2009ib, we measured $m_J=14.88 \pm 0.16$~mag, $(V-J)=1.29 \pm 0.16$~mag, and used $v_{Fe}(50\mathrm{d})=3247 \pm 200$~km$/$s at $50$ days after the explosion. In this way we obtained the distance $D_{\rm SCM,3}=23.16 \pm 3.48$~Mpc.

The three variants of SCM provided slightly different distances. We adopt their mean, $D_{\rm SCM}=23.6 \pm 1.7$~Mpc as the SCM distance to SN~2009ib.

\subsection{Distance to SN~1986L}\label{sec_86L}

SN~1986L was another type II-P supernova that exploded in NGC 1559. It was discovered at a very early phase and its data were also exploited to estimate the host galaxy distance. \citet{Schmidt1994} using an early implementation of the EPM and the dilution factors of \citet{E96} (which were at that time not yet published) estimated the distance as $16 \pm 2$~Mpc. \citet{hamuy_thesis} also used the dilution factors of \citet{E96} and a cross-correlation technique for measuring velocities and obtained the distance of $11.2$~Mpc. According to \citet{hamuy_thesis}, the higher value of  \citet{Schmidt1994} is due to the shallower velocity curve they adopted.

We used photometric measurements in $BV$ bands \citep[M. Phillips, private communication; see also][]{anderson_2014}, and velocities measured from the Doppler-shift of the Fe\,{\sc ii}~$\lambda$5169 line obtained from \citet{Schmidt1994} and \citet{hamuy_thesis}. We interpolated the velocities to the epochs of the photometry using the relations of \citet{takats2012}.  \citet{hamuy_thesis} found that the host galaxy extinction is negligible, therefore we only take into account the galactic extinction in the direction of SN~1986L, which is $E(B-V)_{\rm MW}=0.0258 \pm 0.0003$~mag \citep{reddening_recalib}.

We applied EPM the same way as in the case of SN~2009ib, using the dilution factors of \citet{D05} and photometry in $BV$ band. The calculated parameters can be found in Table~\ref{epm_86l}. We obtained the distance as $D_{\rm 86L}=17.2 \pm 3.0$~Mpc, longer than the previous results of \citet{Schmidt1994} and \citet{hamuy_thesis}, but shorter than the distance determined from the data of SN~2009ib. The difference from the earlier results is mainly due to the dilution factors, which are based on atmospheric models that greatly improved since the first applications of EPM. The lack of $I$ band data usually leads to more uncertain temperature estimates and slightly different distances \citep[see e.g.][]{jones2009}. In the case of SN~2009ib, taking into account only the $BV$ bands would lead to a distance only $4$ per cent lower than with $BVI$ bands.

\begin{table}
 \centering
 \begin{minipage}{84mm}
  \caption{The calculated parameters of SN~1986L needed to apply EPM: the apparent radius ($\theta$), photospheric temperature ($T$), dilution factor ($\zeta(T)$), and the photospheric velocity ($v$), measured from the absorption minimum of Fe\,{\sc ii}~$\lambda5169$ by \citet{Schmidt1994} and \citet{hamuy_thesis} and interpolated to the epochs of the photometry.}
  \label{epm_86l}
  \begin{tabular}{@{}cccccc@{}}
  \hline
  \hline 
 MJD & ${t-t_0 \over 1+z}^a$ & $\theta$ & $T$ & $\zeta(T)$ & $v$ \\  
  & (days) & ($10^8$ ${{\mathrm {km}} \over {\mathrm {Mpc}}}$) & (K) & & (km/s) \\ 
  \hline
46712.4 & 1.8  & 3.495  & 25920 & 0.42  & 30154 \\
46714.3 & 3.7  & 6.569  & 13715 & 0.46  & 18863 \\
46715.3 & 4.7  & 6.537  & 14171 & 0.46  & 16325 \\
46716.2 & 5.6  & 7.549  & 11944 & 0.49  & 14720 \\
46717.3 & 6.7  & 7.495  & 12084 & 0.49  & 13253 \\
46729.3 & 18.6 & 8.516  & 8615  & 0.62  & 7404  \\
46730.4 & 19.7 & 8.993  & 7770  & 0.69  & 7173  \\
46732.4 & 21.7 & 9.289  & 6839  & 0.80  & 6798  \\
46735.2 & 24.5 & 8.713  & 7097  & 0.76  & 6357 \\
46736.2 & 25.5 & 9.954  & 5894  & 0.98  & 6218 \\
46737.4 & 26.7 & 9.484  & 6651  & 0.83  & 6062 \\
46738.4 & 27.7 & 9.428  & 6150  & 0.92  & 5940 \\
46740.3 & 29.6 & 9.223  & 5988  & 0.96  & 5726 \\
46742.3 & 31.6 & 9.072  & 5711  & 1.03  & 5523 \\
46744.3 & 33.6 & 10.447 & 5082  & 1.24  & 5340 \\
46748.2 & 37.4 & 9.836  & 5090  & 1.23  & 5027 \\
46750.3 & 39.5 & 10.535 & 4653  & 1.43  & 4878 \\
46754.3 & 43.4 & 10.832 & 4517  & 1.51  & 4633 \\
46756.3 & 45.4 & 10.699 & 4440  & 1.55  & 4519 \\
\hline
\end{tabular}			  
 \begin{tablenotes}
       \item[a]{$^a$ relative to the epoch of discovery, $t_0=46710.6$~MJD \citep{1986l_discovery}.}
     \end{tablenotes}
\end{minipage}
\end{table}

\subsection{Average Distance}\label{sec_ave}

As mentioned in the beginning of this Section, there have been several previous distance measurements to the host galaxy of SN~2009ib, and those results have a high scatter. 

Most of the measurements used the Tully-Fisher (TF) method \citep{tullyfisher}. Since the TF method was applied by only a few groups, several times either by recalibrating the method or adding newer data, we decided to select only the latest paper from each group \citep{TF1_ngc1559,TF3_ngc1559,TF2_ngc1559}, and then average their results. In this way we estimated the TF distance as $15.7 \pm 1.5$~Mpc. 

The distance to this galaxy was also determined using previous SNe. As discussed in Sect.~\ref{sec_86L},  the EPM method was applied to the Type II-P SN~1986L by \citet{Schmidt1994} and by \citet{hamuy_thesis}, but after reanalysing these data we obtained a distance somewhat higher than the published values. The MLCS2k2 method was used for the Type Ia SN~2005df leading to a distance of $22.0 \pm 2.0$~Mpc \citep{brown2010}. The redshift distance is $14.9 \pm 1.0$~Mpc, however it has significant uncertainties due to the effect of peculiar motions, therefore we exclude it from our analysis. Considering the TF distance, that of SN~2005df, our EPM result for SN~1986L, and the distance to SN~2009ib determined with EPM and SCM (Table \ref{ave_dist}), we conclude the mean of these five results, $D=19.8 \pm 2.8$~Mpc as the distance to NGC~1559.

\begin{table}
 \centering
 \begin{minipage}{84mm}
  \caption{The distances to the host galaxy of SN~2009ib obtained from the literature as well as those determined in this paper. The mean of the values is also shown.}
  \label{ave_dist}
  \begin{tabular}{@{}lccccc@{}}
  \hline
  \hline
Method & $D$ & $\mu$ & References$^a$  \\
& (Mpc) & (mag) & \\
  \hline
Tully-Fisher & 15.7 (1.5) & 30.72 (0.21) & 1,2,3 \\
SN Ia (2005df) & 22.0 (2.0) & 31.72 (0.20) &  4 \\
EPM (1986L) & 17.2 (3.0) & 31.18 (0.37) & this paper \\
EPM (2009ib) & 21.3 (2.2) & 31.64 (0.21) & this paper\\
SCM (2009ib) & 23.6 (1.7) & 31.77 (0.15) & this paper\\
\\
Mean & 19.8 (3.0) & 31.48 (0.31) & \\
\hline				  
\end{tabular}			  
 \begin{tablenotes}
       \item[a]{$^a$ (1) \citet{TF1_ngc1559}, (2) \citet{TF3_ngc1559}, (3) \citet{TF2_ngc1559}, (4) \citet{milne2010}.}
     \end{tablenotes}
\end{minipage}
\end{table}

\section{Bolometric luminosity and Ni mass}\label{sec_bolometric}

\begin{figure}
\includegraphics[width=84mm]{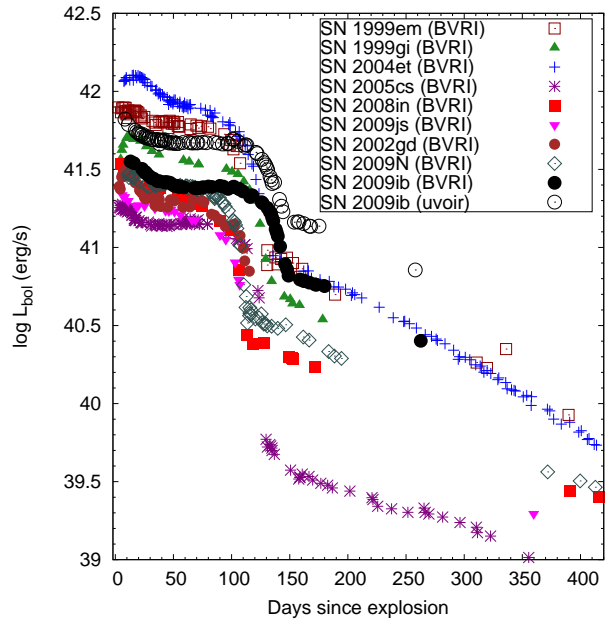}
\caption{Comparison of the evolution of the $BVRI$ quasi-bolometric light curve of SN~2009ib (black filled circles) to those of other SNe~II-P. We also show the $uvoir$ bolometric curve of SN~2009ib (black open circles).}
\label{bollum_comp}
\end{figure}

Using the distance determined in Sect.~\ref{sec_distance}, we can calculate the bolometric luminosity of SN~2009ib. Between days 13 and 104 we have observations covering the wavelength range from $u'$ band up to $H$ band (see Sect.~\ref{sec_photometry}). Using low order polynomials, we interpolated the magnitudes in $u'BRIJH$ bands to the epochs of the observations in $V$ band. After correcting for the reddening, we converted these magnitudes to fluxes, and integrated them using Simpson's rule. The infrared range redwards from $H$ band was estimated assuming that it follows the Rayleigh-Jeans tail of a black body spectrum. After day 104 we lack observations in the NIR bands. At this phase we used the formulae of \citet{maguire_04et}, who estimated the bolometric luminosity from the $R$ and $V$ magnitudes. We find that during the plateau phase this estimation agrees with the luminosity that we calculated, therefore we use them at the later phase, after day 104. The obtained $uvoir$ bolometric luminosity curve is shown in Fig.~\ref{bollum_comp}.

Since the observational coverage for some of the SNe is incomplete, for comparison purposes we computed the luminosities by integrating only the $BVRI$ fluxes. The distances of the comparison SNe can be found in Table~\ref{otherSN}. Fig.~\ref{bollum_comp} shows the comparison of the quasi-bolometric luminosity of SN~2009ib to those of other SNe~II-P. In this comparison we included several SNe that were categorized as subluminous (SNe 2002gd, 2005cs) as well as ``normal'' events (SNe 1999em, 1999gi, 2004et) and ``intermediate'' ones, such as SNe 2008in, 2009N, 2009js. The luminosity of SN~2009ib during its plateau phase is similar to that of SN~2009N. Authors, who examined a larger number of light curves of SNe II \citep[e.g. ][]{anderson_2014, sanders_2014} have found that the luminosity distribution of SNe II-P is continuous, which is reflected in our Figure.

Fig. \ref{bollum_comp} also shows that the plateau phase of SN~2009ib is the longest in the sample. \citet{anderson_2014} found a weak anticorrelation between the plateau length and the absolute brightness including only 15 SNe in the sample. It is clear, that SN~2009ib does not fit in this trend.

The $^{56}$Ni mass produced in the explosion can be estimated from the tail phase luminosity. We used the relation of \citet{hamuy2003} and obtained the value of $0.046 \pm 0.015~{\rm M}_{\sun}$. This value is significantly higher than those of the subluminous and intermediate SNe II-P (Sec.~\ref{sec_comptoSNe}).

\section{Progenitor parameters}\label{sec_physical}

Even though a large number of SNe II have been discovered and studied, we still do not fully understand the exact nature of their progenitors and explosions. In a number of cases direct identification of the progenitor star on archival pre-explosion images have revealed that SNe II-P are originated from red supergiant stars in the mass range $8-16\,{\rm M}_{\sun}$ \citep[eg.][]{smartt_rev}. In other cases, the parameters of the progenitor and its explosion have been estimated via hydrodynamical model calculations \citep[e.g. ][]{litvinova_models, zampieri2003, pumo2011, utrobin99em, Dessart2010}.

In this section we investigate the properties of the progenitor of SN~2009ib first by examining the archival pre-explosion images of its location taken by the HST, then, independently, by hydrodynamical modelling of the SN observables and, finally, we compare the light curve and spectra of SN~2009ib to the non-LTE time-dependent radiative-transfer simulations of \citet{dessart2013}.

\subsection{HST images}\label{sec_hst}

We downloaded the pre-explosion images of the SN site from the HST archive\footnote{\url{https://archive.stsci.edu}}. These images were taken with WFPC2 using the filters $F450W$, $F606W$ and $F814W$ with the total exposure time of $320$~s each, and the pixel scale of $0.0996\arcsec\,{\rm pixel}^{-1}$. The images were obtained on 2001 August 2 (Proposal ID 9042, PI. S. Smartt).

In order to measure the precise position of the supernova, we used post-explosion images taken with VLT$+$NaCo in K$_{\rm s}$ band (PI: Smartt), on 13 August 2009, about 16~days after explosion, downloaded from the ESO archive\footnote{\url{http://archive.eso.org/eso/eso_archive_main.html}}. These images have the pixel scale of $0.05$\arcsec pixel$^{-1}$, the field of view of 56\arcsec $\times$ 56\arcsec, and the total exposure time of $3960$~s. We carried out dark- and flat-field corrections on all 22 images using the standard tasks of {\sc iraf}. The sky was substracted by using off-source sky images taken before and after the on-source images. The sky-substracted frames were then aligned and combined.

In order to achive precise relative astrometry, we geometrically transformed the HST images to the NaCo image using the {\sc geomap} and {\sc geotran} tasks of {\sc iraf}, measuring the positions of 11 point sources that are present on both images. We measure the SN position in the NaCo image with the {\sc iraf} tasks {\sc daofind} and {\sc imexamine}. The error of the position of the SN was calculated from the uncertainties of both the geometric transformations and the position measurements as $160$~mas.

We used the {\sc dolphot} program\footnote{{\sc dolphot} is a stellar photometry package that was adapted from HSTphot for general use by \citet{dolphot}.} to detect and measure sources present in the pre-explosion images. {\sc Dolphot} detected two sources close to the SN position. Both sources were detected at $>5\sigma$ with the filters $F606W$ and $F814W$, but not with filter $F450W$.

\begin{figure*}
\includegraphics[width=\textwidth]{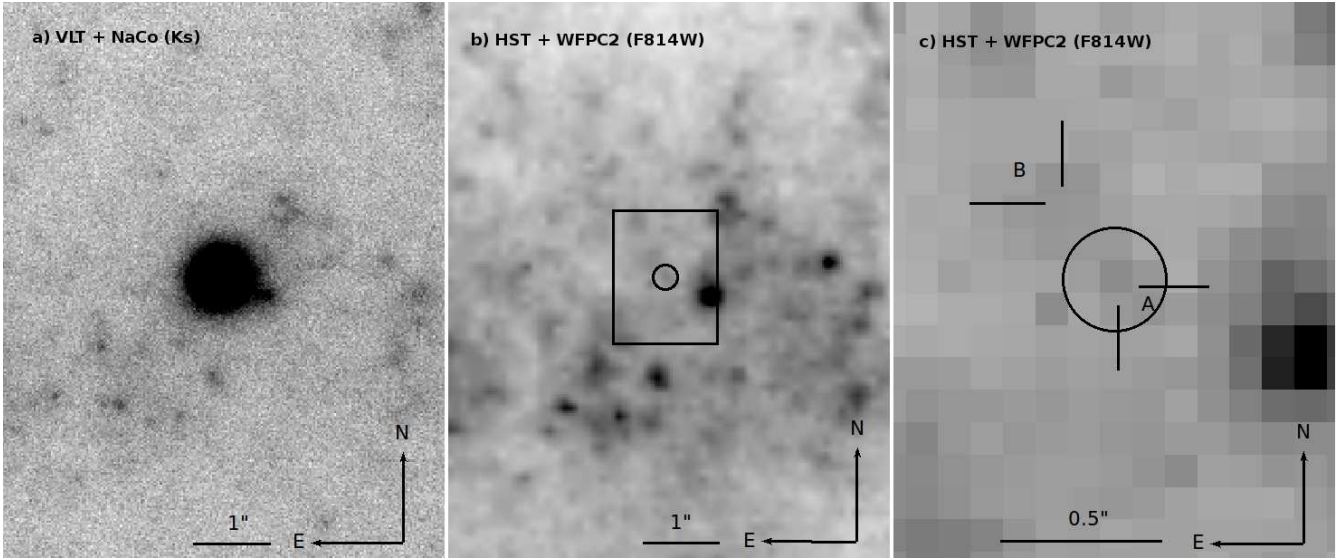}
\caption{(a) The post-explosion image taken with VLT$+$NaCo 15 days after the explosion. (b) A pre-explosion image taken by HST/WFPC2 with the filter F814W. The scale and orientation of the images (a) and (b) are the same. The circle on (b) shows the position of the SN with 1$\sigma$ uncertainty. The area marked with a square is enlarged in (c). The circle in (c) also shows the position of the progenitor, while A and B marks the sources detected with {\sc dolphot} (see text and Table \ref{progenitor_table}). }
\label{prog}
\end{figure*}

Fig.\ref{prog} shows the post-explosion image, as well as the same field in the pre-explosion image (in $F814W$ band). The third image on this figure zooms in to the surroundings of the SN position, with the detected sources marked as ``A'' and ``B''. This figure shows, that source ``A'' is a possible progenitor of the SN. Table \ref{progenitor_table} contains the positions of these sources relative to the SN, as well as their brightness measured by {\sc dolphot}. We converted these magnitudes to the Johnson-Cousins system using the constants provided by \citet{dolphot_constants} and corrected for the reddening (Sect.~\ref{sec_reddening}). The obtained $VI$ values can also be found in Table~\ref{progenitor_table}. 

The spectral type of these sources (assuming they are single stars) were esimated from their $V-I$ colours using Table 4 of \citet{maund_progs}  \citep[see also][]{drilling_spectraltypes}. We found that both sources are likely type G stars. Therefore if source A is the progenitor, it is more yellow, than the the RSG stars that are the usual progenitors of SNe II-P. We used the bolometric corrections of \citet{maund_progs} to calculate the luminosity of source A. We obtained the value of $\log(L/L_{\sun})=5.04 \pm 0.2$~dex.

\citet{anderson_2010} examined the site of SN~1986L, which exploded in the same galaxy as SN~2009ib, and measured the metallicity as $(12+\log({\rm O/H}))=8.51 \pm 0.13$, a solar-like metallicity. Using the spectroscopy image taken with VLT (Sect.\ref{sec_spectroscopy}), we extracted the spectrum of a H\,{\sc ii} region close to SN~2009ib. The distance of this region to SN~2009ib is $17.1\arcsec$, which corresponds to $\sim 1.6$~kpc. It is closer than SN~1986L, the distance of which is $65.5\arcsec$. From the spectrum of the H\,{\sc ii} region we measured the ratios of $[$N\,{\sc ii}$] \lambda$6583$/$H$\alpha$ (N2) and ($[$O\,{\sc iii}$] \lambda$5007$/$H$\beta$)$/$N2 (O3N2), and following the procedure of \citet{anderson_2010} and using the relations of \citet{pettini_metallicity} we obtain the $N2$ metallicity $(12+\log({\rm O/H}))_{\rm N2}=8.44 \pm 0.21$ and the $O3N2$ metallicity $(12+\log({\rm O/H}))_{\rm O3N2}=8.73 \pm 0.17$. This value is similar to that of \citet{anderson_2010} and close to the solar metallicity.

\begin{figure}
\includegraphics[width=84mm]{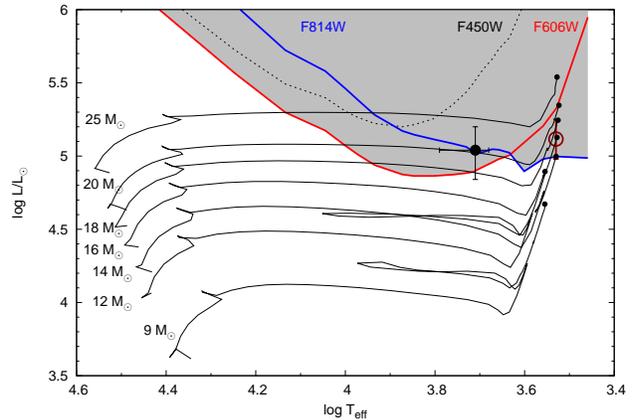}
\caption{A HR diagram showing the  evolution tracks of massive stars with solar-like metallicity from the {\sc stars} code. The filled circle marks the position of the yellow source ``A''. It lies on the evolution track of an $M_{\rm ZAMS}=20\,{\rm M}_{\sun}$ star. The empty circle marks the position of an RSG progenitor which was calculated based on the  assumption that only the $F814W$ band detection belongs to the progenitor (see text). This star lies on the $M_{\rm ZAMS}=16\,{\rm M}_{\sun}$ track. The bolometric luminosities calculated from the measured magnitude upper limits are plotted as a function of the effective temperature with black dashed ($F450W$), red ($F606W$) and blue ($F814W$) continuous lines. This figure shows, that if the progenitor were an RSG star more massive than $\sim 14~{\rm M}_{\sun}$, we would have been able to detect it with the filter $F814W$ and if more than $\sim 20~{\rm M}_{\sun}$, both with $F606W$ and $F814W$.}
\label{masslimit}
\end{figure}

In Fig.~\ref{masslimit} we plotted the position of source A on a Herzsprung-Russel (HR) diagram together with stellar evolution tracks from the {\sc stars} code \citep[see][and the references therein]{eldridge_stars_code} with solar metallicity. This diagram shows that source A is on the evolution track of the $20\,{\rm M}_{\sun}$ star. 

 In a few cases there have been a yellow source found at the SN postition in pre-explosion images. \citet{li_04et} found the progenitor of SN~2004et to be a yellow supergiant with a ZAMS mass of $15^{+5}_{-2}\,{\rm M}_{\sun}$, but \citet{smartt2009} argued that that source is not a single star. \citet{eliasrosa_2008cn} also found a yellow source at the position of the progenitor of SN~2008cn. Other than having a yellow supergiant with a ZAMS mass of $15 \pm 2\,{\rm M}_{\sun}$ as a progenitor, they offer alternative explanations for the yellow colour of the source. They consider the possibility that it is not a single star, instead a blending of two or more star. Examining model stellar spectra, they showed that the yellow colour can result e.g. from a blend of a type B0 with $M_{ini} \approx 40\,{\rm M}_{\sun}$ and a type M1 star with $M_{ini} \approx 15\,{\rm M}_{\sun}$.  In fact, by taking very late time images of this SN with HST, \citet{maund_progenitorsrevisited} showed that the pre-explosion yellow source contained a blue star that remained present after the explosion, and concluded that the progenitor was most likely an RSG star with $M_{ini} \approx 16\,{\rm M}_{\sun}$. In a similar case, \citet{Fraser2010} and \citet{elias-rosa_2009kr} both arrived to the conclusion that the progenitor of SN~2009kr was a yellow supergiant star, with initial mass of $18-24\,{\rm M}_{\sun}$ and $15^{+5}_{-4}\,{\rm M}_{\sun}$, respectively. Revisiting this SN at late times, \citet{maund_progenitorsrevisited} found that the yellow source is probably a small compact cluster, not an single source.

In the case of SN~2009ib, we can examine a scenario similar to that of SN~2008cn. We can assume, that in the $F814W$ filter we detect the RSG progenitor, but due to its red color we do not see it in the $F606W$ band, where the detection is due to a blue star, unrelated to the SN. Assuming that the RSG star has a temperature of $\sim 3400\,{\rm K}$, we determined the luminosity of the RSG star as $\log(L/L_{\sun})=5.12 \pm 0.14$~dex, following the same procedure as before. We show the position of this source on the HR diagram in Figure~\ref{masslimit}. It shows that this RSG star would have had an initial mass of $16 \pm 2\,{\rm M_{\sun}}$.

\begin{table}
 \centering
 \begin{minipage}{84mm}
  \caption{Position and brightness of the sources found by {\sc dolphot} in the pre-explosion HST images close to the SN position. }
  \label{progenitor_table}
  \begin{tabular}{@{}lcc@{}}
  \hline
  \hline
& Source A & Source B \\
\hline
$\Delta \alpha$/$\Delta \delta$ (mas)$^a$ & 4/50 & 20/310 \\
F606W (mag) &  23.91 (0.11) & 24.05 (0.17) \\
F814W (mag) & 23.20 (0.12) & 23.12 (0.13) \\
$V^b$ (mag) & 23.79 (0.18) & 24.02 (0.24) \\
$I^b$ (mag) & 22.94 (0.17) & 22.83 (0.21) \\
Spectral type$^c$ & G0-G1 (F6-G6) & G8-K0 (G1-K3)  \\
\hline				  
\end{tabular}			  
 \begin{tablenotes}
       \item[a]{$^a$ distance to the SN position.}
       \item[b]{$^b$ reddening corrected values.}
       \item[c]{$^c$ in parenthesis we give the spectral type range estimated considering the uncertainties of the $VI$ magnitudes.}
     \end{tablenotes}
\end{minipage}
\end{table}

We should also consider, that source A is not the progenitor, which, instead, is too faint to be visible in the pre-explosion images. Therefore we estimated the detection limits by running artificial star experiments using {\sc dolphot}. We found that the upper limits at the position of the progenitor are $23.87$~mag for $F450W$, $24.30$~mag for $F606W$ and $23.25$~mag for $F814W$. Correcting for the reddening (Sect.\ref{sec_reddening})  and taking into account the distance to the SN (Sect.\ref{sec_distance}), these values correspond to the absolute magnitudes $M_{F450W} \gtrsim -8.12$~mag, $M_{F606W} \gtrsim -7.55$~mag and $M_{F814W} \gtrsim -8.62$~mag.

We used the colour and bolometric corrections of \citet{maund_progs} to convert the measured upper limits to bolometric magnitudes. Since in this scenario we have no precise colour information of the progenitor, we calculated and plotted the lower limit of the bolometric luminosity as the function of the effective temperature and plotted them on Fig.~\ref{masslimit}.
It shows that assuming that the progenior was an RSG star, we should have seen it in filter $F814W$, if its initial zero-age main sequence mass exceeded $14~{\rm M}_{\sun}$ and both in filters $F606W$ and $F814W$ if its mass was $M_{ZAMS} > 20\,{\rm M}_{\sun}$. Note, however, stellar evolution track calculations have several uncertainties. The final luminosity of a star with a given mass depends heavily on the treatment of the mixing processes, e.g. a lower overshooting parameter leads to lower luminosity. As \citet{eldridge_tout_stars} discussed, by turning the convective overshooting off in the STARS code, the final luminosity of a star with a certain initial mass increases. Therefore the upper limit of the progenitor mass in the case of SN~2009ib would be $2-3\,{\rm M}_{\sun}$ higher without overshooting. In addition, \citet{walmswell_eldridge} discussed that during their lives red supergiant stars (which were found to be the progenitors of several SNe II-P) produce dust that is destroyed in the SN explosion, therefore the extinction before the explosion can be significantly higher than the one estimated after the explosion, also leading to higher upper mass limits. Both of these uncertainties point toward a higher mass limit, therefore it is reasonable to consider that the upper mass limit of an RSG star too faint to be undetected in the pre-explosion images would be between $14-17\,{\rm M}_{\sun}$

In summary, there are several possible interpretations of the pre-explosion HST images. These scenarios imply that the progenitor of SN~2009ib had an initial mass between $14-20\,{\rm M}_{\sun}$. However, high resolution, late time HST images are necessary to be able to rule out some of these possibilities, similarly as was done by \citet{maund_progenitorsrevisited}.

\subsection{Hydrodynamical modelling}\label{sec_hydro}

We constrain the main physical properties of SN 2009ib at the explosion (namely the ejected mass, the progenitor radius and the explosion energy) through the hydrodynamical modeling of the SN observables (i.e. bolometric light curve, evolution of line velocities and continuum temperature at the photosphere). 

We use the same approach adopted for other SNe  \citep[e.g. 2007od, 2009E, 2009N, 2012A, and 2012aw; see ][respectively]{inserra2011, pastorello_09E, takats2014, tomasella_12A, dallora_12aw}, in which a simultaneous $\chi^{2}$ fit of the observables against model calculations is performed and two codes are employed for computing the models. The first one is the semi-analytic code described in detail by \citet{zampieri2003} which is used to perform a preparatory study aimed at individuating the approximate location in the parameter space describing the SN progenitor at the explosion. The results of such study are utilised to guide the more realistic, but time consuming model calculations performed with the second code which is the general-relativistic, radiation-hydrodynamics Lagrangian code presented in \citet{pumo2010} and \citet{pumo2011}. Its main features are \citep[cfr. also ][]{pumo2013}: a) a fully implicit Lagrangian approach to the solution of the coupled non-linear finite difference system of relativistic radiation-hydro equations, b) an accurate treatment of radiative transfer in all regimes (from the one in which the ejected material is optically thick up to when it is completely transparent), and c) a description of the evolution of ejected material which takes into account both the gravitational effects of the compact remnant and the heating effects linked to the decays of the radioactive isotopes synthesized during the SN explosion.

\begin{figure}
\includegraphics[angle=270,width=84mm]{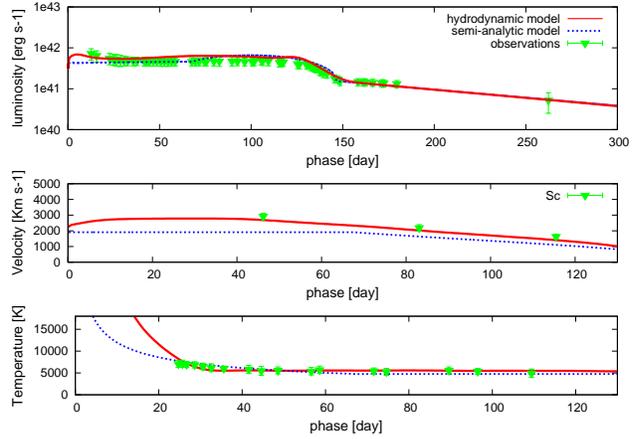}
\caption{Comparison of the evolution of the main observables of SN 2009ib with the best-fit models computed with the general-relativistic, radiation-hydrodynamics code (total energy $0.55$ foe, pre-explosion radius $2.8 \times 10^{13}$ cm, envelope mass $15\,{\rm M_{\sun}}$). Top, middle, and bottom panels show the bolometric light curve, the photospheric velocity, and the photospheric temperature as a function of time. To better estimate the photosphere velocity from observations, we use the minima of the profile of the Sc\,{\sc ii} lines which are considered good tracer of the photosphere velocity in Type II SNe. For the sake of completeness, the best-fit model computed with the semi-analytic code (total energy $0.4$ foe, initial radius $2.4 \times 10^{13}$ cm, envelope mass $12.4\,{\rm M}_{\sun}$) is also shown.}
\label{fighydro}
\end{figure}

Based on the adopted explosion epoch (MJD= 55041.3; Sect.~\ref{sec_explepoch}), bolometric luminosity and nickel mass ($M(^{56}{\rm Ni})=0.046\,{\rm M_{\sun}}$; Sect.~\ref{sec_bolometric}), the best fit gives values of total (kinetic plus thermal) energy of $0.55$ foe, progenitor radius of $2.8 \times 10^{13}$ cm ($\sim 400\,{\rm R_{\sun}}$) and envelope mass of $15\,{\rm M_{\sun}}$ (see Fig.~\ref{fighydro}). The estimated uncertainty on the modelling parameters are about 15 per cent.

Adding the mass of the compact remnant ($\sim 1.5-2.0\,{\rm M_{\sun}}$) to that of the ejected material, we estimate the mass of the progenitor of SN 2009ib at the explosion as $16.5-17\,{\rm M_{\sun}}$, which can be even higher if we consider the mass that the progenitor may have lost during its life. This estimate is consistent with the mass range of RSG precursors, and somewhat higher than the upper limit of the progenitor mass determined from the pre-explosion images (Sect. \ref{sec_hst}). This is in agreement with results in the literature that found that progenitor mass estimate from hydrodynamical modelling are in general larger than those from direct detections \citep[see e.g.][]{smartt_rev, maguire_04et, tomasella_12A}.

\subsection{Comparison to non-LTE models}

We compared the observational properties of SN~2009ib to the non-LTE time-dependent radiative-transfer simulations of \citet{dessart2013}. They used {\sc mesa star} \citep{paxton_mesa,paxton_mesa2} to evolve a $15\,{\rm M}_{\sun}$ main-sequence star. Changing parameters such as mixing length, overshoot, rotation and metallicity, they created a grid of pre-SN model stars. Then they generated a piston-driven explosion with {\sc v1d} \citep{livne_v1d,dessart_2010b} and evolved the SN with {\sc cmfgen} \citep{hillier_cmfgen,Dessart2005a,dessart2008,hillier_2012} until late into its nebular phase \citep[for details see][]{dessart2013}. The model properties can be found in Table 1 of \citet{dessart2013}. They compared the light curve and spectra of SN~1999em to the models, and found the only model able to reproduce the colour evolution is the one called m15mlt3, the one where the RSG progenitor have significantly smaller pre-explosion radius ($\sim 500\,{\rm R}_{\sun}$) than the rest. 

\begin{figure}
\includegraphics[width=84mm]{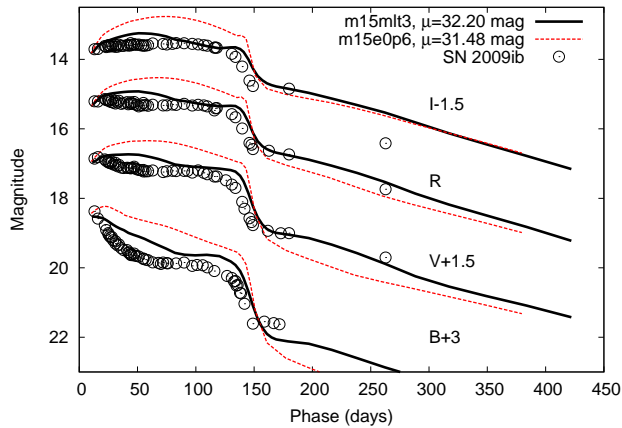} 
\caption{Comparison of the light curve of SN~2009ib to the models m15mlt3 and m15e0p6 of \citet{dessart2013}. The m15e0p6 model is scaled to the distance of the SN and matches its brigthness well, while the brightness of the model m15e0p6 is higher, therefore we applied an additional vertical shift of $0.88$ mag for better comparison. }
\label{dessartmodel_lc}
\end{figure}

\begin{figure*}
\includegraphics[width=\textwidth]{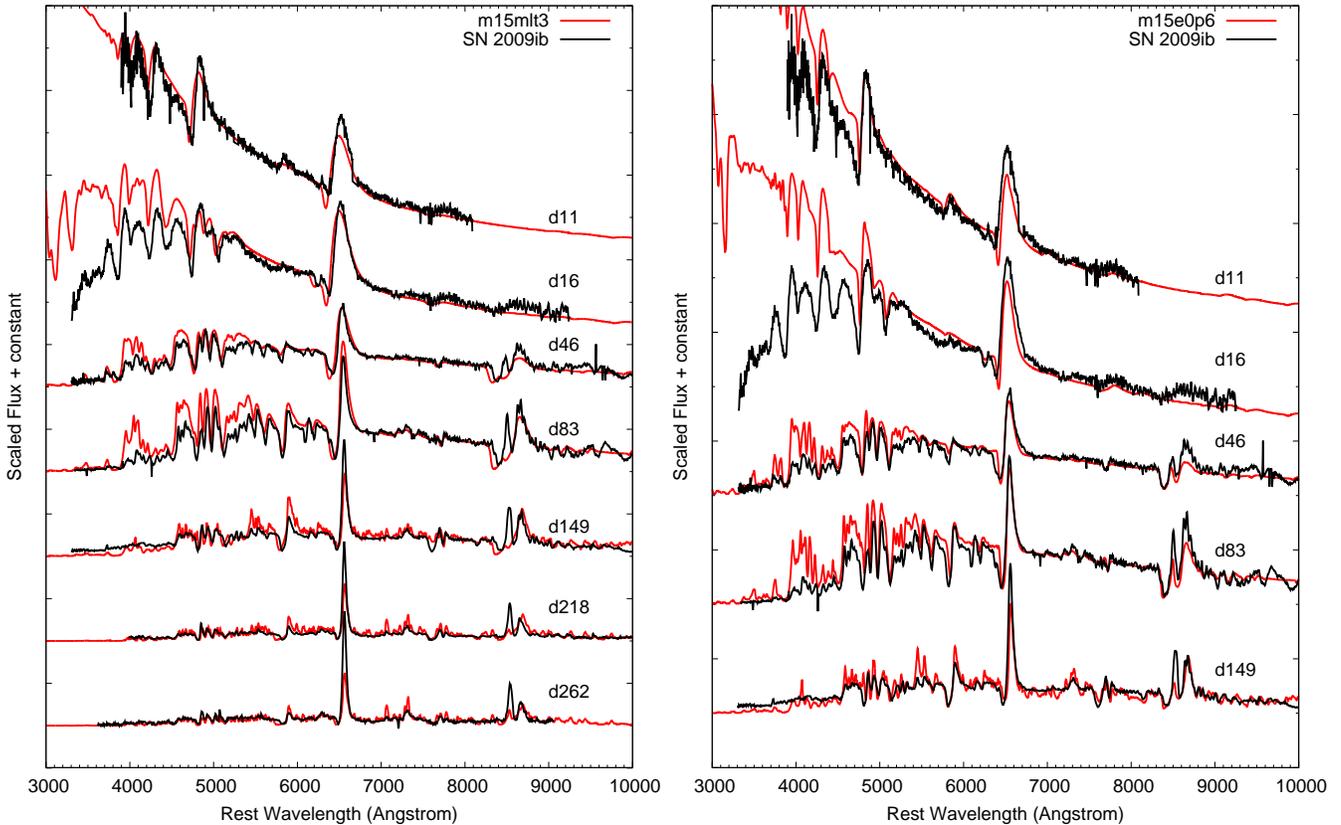}
\caption{Comparison of the optical spectra of SN~2009ib with those of the models m15mlt3 (left) and m15e0p6 (right) of \citet{dessart2013}. There is an obvious colour mismatch in the case of the day 16 spectrum with both models. The overall fit is not bad, though the velocities and the strength of the features do not match everywhere.}
\label{dessartmodel_sp}
\end{figure*}

We compared the properties of SN~2009ib to two of the models: to m15mlt3 mentioned above, and to the model m15e0p6, which had lower ejecta kinetic energy than the baseline model ($0.6$ foe instead of $1.2$ foe). Fig.~\ref{dessartmodel_lc} compares the light curve of SN~2009ib to the two models, while Fig.~\ref{dessartmodel_sp} shows the spectral comparison. It is clear from these comparisons, that while the model m15mlt3 fits better the colour evolution and the length of the plateau phase, the model m15e0p6 reproduces better the brightness and the spectral evolution of the SN. Consequently, the best-fitting model might be the combination of these two, a model with a smaller progenitor radius and small ejecta kinetic energy. The progenitor radius of the m15mlt3 model and the kinetic energy of the m15e0p6 model are close to the values we obtained from the hydrodynamical modelling. The hydrogen envelope mass for both models is $10.2\,{\rm M}_{\sun}$. \citet{dessart2013} notes, that the ejecta mass depends also on the He core mass, which only can be estimated via nebular phase modelling, which goes beyond the scope of this paper. There is a broad agreement between the our early nebular phase spectra and the models, therefore the best-fitting model should not be very far from those we examined here, but a more careful comparison to a broader grid of models is necessary to draw more accurate conclusions.

\subsection{Comparison to other SNe II-P}\label{sec_comptoSNe}

The brightness of SN~2009ib places it among the ``intermediate'' luminosity SNe II-P such as SNe~2008in and 2009N, however it also shows many differences. SN~2009ib has an unusually long plateau,  which can be explained by its relatively massive hydrogen-rich envelope. The hydrodynamical modelling revealed an ejecta mass of $\sim 15\,{\rm M}_{\sun}$, significantly higher than those of SN~2009N \citep[$11.5\,{\rm M}_{\sun}$][]{takats2014}, SN~2008in \citep[$13.5\,{\rm M}_{\sun}$][]{spiro_2014}.

The estimated amount of $^{56}$Ni mass ($0.046\,{\rm M}_{\sun}$) is closer to those of the normal SNe II-P, e.g. 2004et \citep[$0.056\,{\rm M}_{\sun}$,][]{maguire_04et} or 1999em \citep[$0.036\,{\rm M}_{\sun}$,][]{utrobin99em}, and significantly higher than those of the subluminous SNe such as SN~2005cs \citep[$0.003-0.004\,{\rm M}_{\sun}$,][]{pastorello05csII} and intermediate luminosity SNe \citep[e.g. $0.020\,{\rm M}_{\sun}$ for SN~2009N,][]{takats2014}. The explosion energy ($0.55$~foe) on the other hand is similar to those of the intermediate luminosity SNe (e.g.  $0.48$~foe for SN~2009N) and between those of the normal ones \citep[e.g. $1.3$~foe for SN~1999em,][]{zampieri_sne_conf}, and those of the subluminous ones (e.g. $0.2-0.4$~foe for SN~2005cs).

\begin{figure*}
\includegraphics[width=\textwidth]{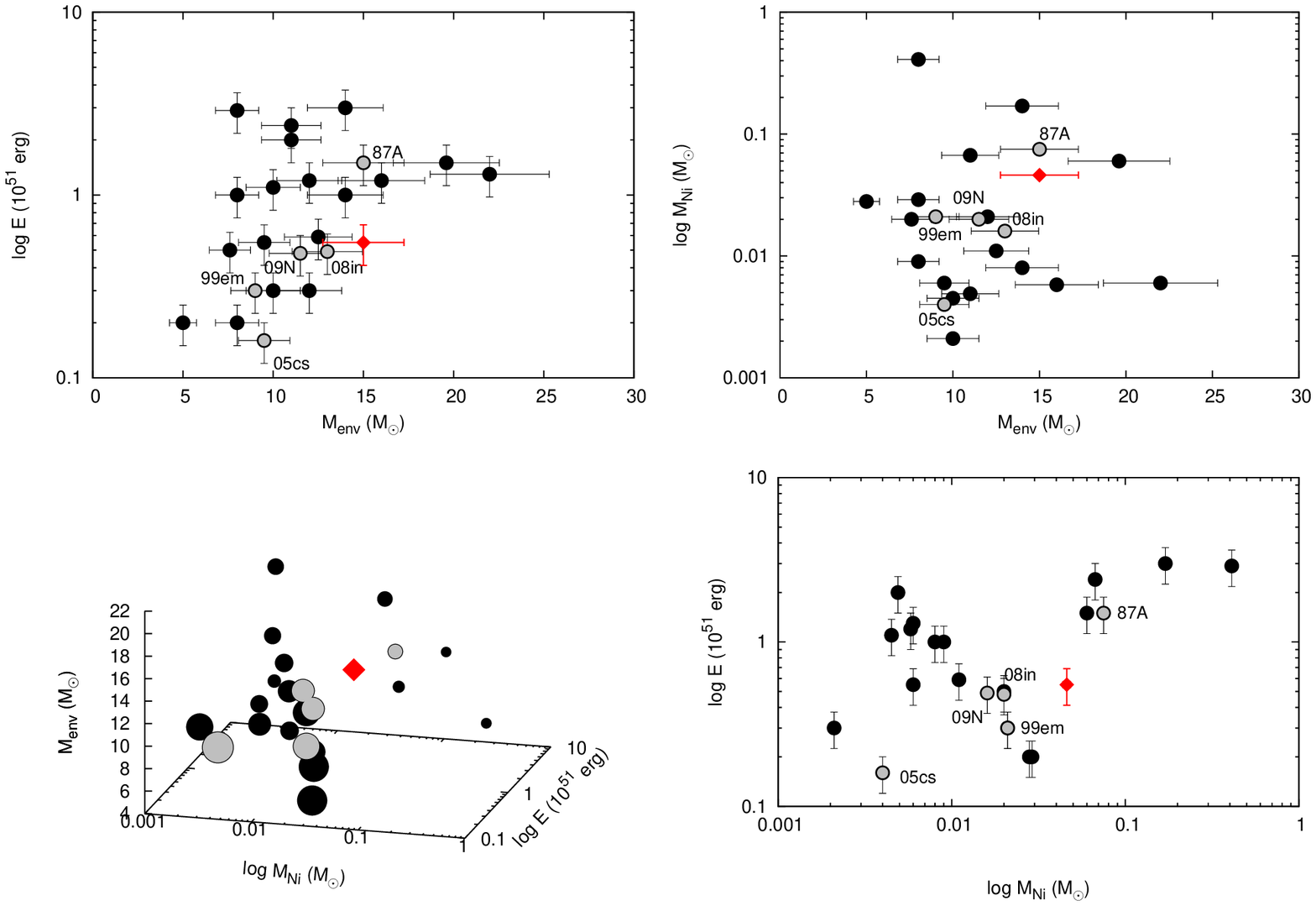}
\caption{Comparison of the explosion energy ($E$), nickel mass ($M_{\rm Ni}$) and ejected envelope mass ($M_{\rm env}$) of SN~2009ib (red diamond) estimated via hydrodynamical modelling to those of other SNe II-P. The sources of the data used for comparison are \citet{zampieri_sne_conf},  \citet{inserra2011}, \citet{inserra2012}, \citet{inserra2013}, \citet{tomasella_12A}, \citet{takats2014}, \citet{dallora_12aw} and \citet{spiro_2014}. In the bottom left figure the size of the points are inversely proportional to $\log(E)$, to better show their position along that axis. }
\label{physical_comp}
\end{figure*}

In Fig.~\ref{physical_comp} we compare the physical parameters of SN~2009ib and its progenitor to those of other SNe II-P. The majority of the data we used here were published in \citet{zampieri_sne_conf} who carried out hydrodynamical modelling to estimate the parameters of SNe and compared them using similar plots as Fig.~\ref{physical_comp}. We also included in the comparison 
SNe~2007od \citep{inserra2011}, 2009N \citep{takats2014}, 2009bw \citep{inserra2012}, 1995ad, 2009dd, and 2010aj \citep{inserra2013}, 2012A \citep{tomasella_12A}, 2012aw \citep{dallora_12aw}, 2005cs and 2008in \citep{spiro_2014}. The physical parameters of these objects were estimated with the same hydrodynamical modelling technique that we use in this paper (Sect.~\ref{sec_hydro}). In the Figure we highlighted the subluminous SN~2005cs, the normal SN~1999em and the peculiar SN~1987A, which objects are often used to represent their subgroups, as well as SNe~2008in and 2009N, which have the same brightness as SN~2009ib.

Fig.~\ref{physical_comp} confirms the intermediate nature of SN~2009ib, which falls mostly in the middle of the examined parameter space. Its nickel mass is among the higher ones in this sample. Its envelope mass is the closest to that of the peculiar SN 1987A, which emerged from a blue supergiant progenitor and has different observational properties than the majority of SNe II-P. The luminosity as well as the explosion energy of SN~2009ib is very close to those of SNe~2008in and 2009N, but the nickel mass and hydrogen envelope mass are higher than those of the other two, which can explain the longer plateau phase.

\section{Summary}\label{sec_summary}

In this paper we present data and analysis of SN~2009ib in NGC~1559. This object lies between the luminosity ranges of subluminous and normal SNe~II-P, belonging to the growing sample of the so-called ``intermediate'' luminosity objects, such as SNe~2008in and 2009N. However, this object shows significant differences from those, further blurring the line between the sub-groups of SNe II-P. 

Among the most important observational properies of SN~2009ib are its moderate bolometric luminosity, unusually long plateau (lasting 120 days after explosion) and moderate expansion velocities. The spectra is very similar to that of the subluminous SN~2002gd, but shows differences from those of other subluminous and intermediate luminosity SNe. Comparing the NIR spectra to those of other objects, we find that the Sr\,{\sc ii} lines weaker and the C\,{\sc i} lines are significantly stronger than usual (noting, however, that the available sample is quite limited). 

We estimated the distance to SN~2009ib using both the expanding photosphere method and the standardized candle method. We also determined the distance to SN~1986L, another SN II-P that exploded in the same galaxy. Combining these results with distances from the literature we obtained the value of $D=19.8 \pm 3.0$~Mpc ($\mu=31.48 \pm 0.31$~mag). Using this distance, we constructed the bolometric light curve of the SN, and from the tail luminosity we estimated the mass of the $^{56}$Ni produced in the explosion as $0.046 \pm 0.015\,{\rm M}_{\sun}$, a value comparable to those of normal SNe~II-P, such as SN~2004et.

We examined the pre-explosion images of the galaxy taken by the HST, searching for the progenitor. 
 We found two sources close to the SN, one of them coinciding with its position. We detected this source in the filters $F606W$ and $F814W$. Assuming it is a single star, the yellow colour and the luminosity of the source would indicate a type G1 star with $M_{\rm ZAMS} \approx 20\,{\rm M}_{\sun}$. It is also possible, however, that it is a blend of two or more stars, and the yellow colour is due to the superposition of the light from those. Assuming that in the $F606W$ band an accompanying blue star is detected while the $F814W$ band detection belongs to the progenior and this progenitor is an RSG star, we obtained the mass estimate $M_{\rm ZAMS} \approx 16\,{\rm M}_{\sun}$. In addition, we examined the possibility that the progenitor of the SN is too faint to appear in HST images. Assuming that it exploded as a RSG star, we found that the upper limit of this mass to be $14-17\,{\rm M}_{\sun}$.

Using hydrodynamical modelling, we estimated the progenitor parameters at the explosion, and found the ejecta mass $15$~M$_{\sun}$, the total explosion energy $0.55$~foe and the inital radius as $2.8 \times 10^{13}$~cm. Adding the mass of a compact remnant, the estimated progenitor mass is $16.5-17\,{\rm M_{\sun}}$. 

We compared the light curve and spectra of SN~2009ib to the non-LTE time-dependent radiative-transfer simulations of \citep{dessart2013}, where the main sequence mass of the model star was $15\,{\rm M}_{\sun}$. We found that the models with reduced inital radius match reasonably the colour evolution, plateau length and spectral evolution of the SN, while we need the model with smaller kinetic energy to match the absolute brightness.

In summary, we found that SN~2009ib may have emerged from a red supergiant star with a main sequence mass of $14-20\,{\rm M}_{\sun}$ and an initial radius of $\sim 400-500\,{\rm M}_{\sun}$. Its explosion had moderate energy of $\sim 0.6$~foe, resulting in a SN with intermediate luminosity and expansion velocity, but somewhat elevated amount of $^{56}$Ni and an unusually long plateau phase.

\section*{Acknowledgements}

We thank the referee for the useful comments that helped to improve this manuscript. We thanks Luc Dessart for sending us his models in digital format and for his comments;  Mark Phillips for sending us the photometric data of SN~1986L; and Giorgos Leloudas for observing SN~2009ib.

K.T. was supported by the Gemini-CONICYT Fund, allocated to the project N$^o$ 32110024 and by CONICYT through the FONDECYT grant 3150473. Support for K.T., G.P., F.B. and M.H. is provided by the Ministry of Economy, Development, and Tourism's Millennium Science Initiative through grant IC12009, awarded to the Millennium Institute of Astrophysics, MAS.
We acknowledge the TriGrid VL project and the INAF-Astronomical Observatory of Padua for the use of computer facilities. M.L.P.~acknowledges financial support from CSFNSM and from the PRIN-INAF 2011 ``Transient Universe: from ESO Large to PESSTO'' (P.I.~S.~Benetti). S.B., N.E.R. and E.C. are also partially supported by the same PRIN-INAF. N.E.R. acknowledges the support from the European Union Seventh Framework Programme (FP7/2007-2013) under grant agreement n. 267251 "Astronomy Fellowships in Italy"  (AstroFIt).  M. F. received support from the European Union FP7 programme through ERC grant number 320360. S.J.S. acknowledges funding from the European Research Council under the European Union's Seventh Framework Programme (FP7/2007-2013)/ERC Grant agreement n$^{\rm o}$ [291222] and STFC grants ST/I001123/1 and ST/L000709/1. M. D. S. gratefully acknowledges generous support provided by the Danish Agency for Science and Technology and Innovation realized through a Sapere Aude Level 2 grant.

Data of this work have been taken in the framework of the European supernova collaboration involved in ESO-NTT large programme 184.D-1140 led by Stefano Benetti. Partially based on observations made with the REM Telescope, INAF Chile. We wish to thank the REM team for technical support, and in particular Dino Fugazza, for their help in setting-up the observations.
Based on observations obtained at the Gemini Observatory, which is operated by the 
Association of Universities for Research in Astronomy, Inc., under a cooperative agreement 
with the NSF on behalf of the Gemini partnership: the National Science Foundation 
(United States), the National Research Council (Canada), CONICYT (Chile), the Australian 
Research Council (Australia), Minist\'{e}rio da Ci\^{e}ncia, Tecnologia e Inova\c{c}\~{a}o 
(Brazil) and Ministerio de Ciencia, Tecnolog\'{i}a e Innovaci\'{o}n Productiva (Argentina). Data were obtained under the Gemini programs GS-2009B-Q-40 and GS-2009B-Q-67. This research is based in part on observations made with the SMARTS Consortium 1.3 m telescope and the Prompt Telescopes located at Cerro Tololo Inter-American Observatory (CTIO), Chile; the Very Large Telescope located at Paranal Observatory under the programmes 084.D-0261 and 083.D-0131. Based on observations made with the NASA/ESA Hubble Space Telescope, obtained from the archive at the Space Telescope Science Institute, which is operated by the Association of Universities for Research in Astronomy, Inc., under NASA contract NAS 5-26555. These observations are associated with program \#9042.

This research has made use of the NASA/IPAC Extragalactic Database, the HyperLeda data base, NASA’s Astrophysics Data System. The availability of these services is gratefully acknowledged.

\bibliographystyle{mn2e}
\bibliography{./refs}

\begin{thebibliography}{}

\bibitem[\protect\citeauthoryear{{Anderson}, {Covarrubias}, {James}, {Hamuy} \&
  {Habergham}}{{Anderson} et~al.}{2010}]{anderson_2010}
{Anderson} J.~P.,  {Covarrubias} R.~A.,  {James} P.~A.,  {Hamuy} M.,
  {Habergham} S.~M.,  2010, MNRAS, 407, 2660

\bibitem[\protect\citeauthoryear{{Anderson} et~al.,}{{Anderson}
  et~al.}{2014}]{anderson_2014}
{Anderson} J.~P.  et~al., 2014, ApJ, 786, 67

\bibitem[\protect\citeauthoryear{{Benetti} et~al.,}{{Benetti}
  et~al.}{2001}]{benetti_1997D}
{Benetti} S.  et~al., 2001, MNRAS, 322, 361

\bibitem[\protect\citeauthoryear{{Blondin} \& {Tonry}}{{Blondin} \&
  {Tonry}}{2007}]{SNID}
{Blondin} S.,  {Tonry} J.~L.,  2007, ApJ, 666, 1024

\bibitem[\protect\citeauthoryear{{Dall'Ora} et~al.,}{{Dall'Ora}
  et~al.}{2014}]{dallora_12aw}
{Dall'Ora} M.  et~al., 2014, ApJ, 787, 139

\bibitem[\protect\citeauthoryear{Dessart \& Hillier}{Dessart \&
  Hillier}{2005}]{Dessart2005a}
Dessart L.,  Hillier D.~J.,  2005, A\&A, 437, 667

\bibitem[\protect\citeauthoryear{{Dessart} \& {Hillier}}{{Dessart} \&
  {Hillier}}{2005}]{D05}
{Dessart} L.,  {Hillier} D.~J.,  2005, A\&A, 439, 671

\bibitem[\protect\citeauthoryear{Dessart, Livne \& Waldman}{Dessart
  et~al.}{2010}]{Dessart2010}
Dessart L.,  Livne E.,    Waldman R.,  2010, MNRAS, 408, 827

\bibitem[\protect\citeauthoryear{{Dessart}, {Livne} \& {Waldman}}{{Dessart}
  et~al.}{2010}]{dessart_2010b}
{Dessart} L.,  {Livne} E.,    {Waldman} R.,  2010, MNRAS, 405, 2113

\bibitem[\protect\citeauthoryear{{Dessart}, {Hillier}, {Waldman} \&
  {Livne}}{{Dessart} et~al.}{2013}]{dessart2013}
{Dessart} L.,  {Hillier} D.~J.,  {Waldman} R.,    {Livne} E.,  2013, MNRAS,
  433, 1745

\bibitem[\protect\citeauthoryear{{Dessart} et~al.,}{{Dessart}
  et~al.}{2008}]{dessart2008}
{Dessart} L.  et~al., 2008, ApJ, 675, 644

\bibitem[\protect\citeauthoryear{{Dolphin}}{{Dolphin}}{2000}]{dolphot}
{Dolphin} A.~E.,  2000, PASP, 112, 1383

\bibitem[\protect\citeauthoryear{{Dolphin}}{{Dolphin}}{2009}]{dolphot_constants}
{Dolphin} A.~E.,  2009, PASP, 121, 655

\bibitem[\protect\citeauthoryear{{Drilling} \& {Landolt}}{{Drilling} \&
  {Landolt}}{2000}]{drilling_spectraltypes}
{Drilling} J.~S.,  {Landolt} A.~U.,  2000, {Normal Stars}.
p.~381

\bibitem[\protect\citeauthoryear{Eastman, Schmidt \& Kirshner}{Eastman
  et~al.}{1996}]{E96}
Eastman R.~G.,  Schmidt B.~P.,    Kirshner R.,  1996, ApJ, 466, 911

\bibitem[\protect\citeauthoryear{{Eldridge} \& {Tout}}{{Eldridge} \&
  {Tout}}{2004}]{eldridge_tout_stars}
{Eldridge} J.~J.,  {Tout} C.~A.,  2004, MNRAS, 353, 87

\bibitem[\protect\citeauthoryear{{Eldridge}, {Izzard} \& {Tout}}{{Eldridge}
  et~al.}{2008}]{eldridge_stars_code}
{Eldridge} J.~J.,  {Izzard} R.~G.,    {Tout} C.~A.,  2008, MNRAS, 384, 1109

\bibitem[\protect\citeauthoryear{{Elias-Rosa} et~al.,}{{Elias-Rosa}
  et~al.}{2009}]{eliasrosa_2008cn}
{Elias-Rosa} N.  et~al., 2009, ApJ, 706, 1174

\bibitem[\protect\citeauthoryear{{Elias-Rosa} et~al.,}{{Elias-Rosa}
  et~al.}{2010}]{elias-rosa_2009kr}
{Elias-Rosa} N.  et~al., 2010, ApJL, 714, L254

\bibitem[\protect\citeauthoryear{{Evans}, {McNaught}, {Cragg} \&
  {Thompson}}{{Evans} et~al.}{1986}]{1986l_discovery}
{Evans} R.,  {McNaught} R.,  {Cragg} T.,    {Thompson} G.,  1986, IAU Circ.,
  4260, 1

\bibitem[\protect\citeauthoryear{{Evans}, {Overbeek} \& {Thompson}}{{Evans}
  et~al.}{1984}]{1984j_discovery}
{Evans} R.,  {Overbeek} D.,    {Thompson} G.,  1984, IAU Circ., 3963, 1

\bibitem[\protect\citeauthoryear{Fraser et~al.,}{Fraser
  et~al.}{2011}]{fraser09md}
Fraser M.  et~al., 2011, MNRAS, 417, 1417

\bibitem[\protect\citeauthoryear{Fraser et~al.,}{Fraser
  et~al.}{2010}]{Fraser2010}
Fraser M.  et~al., 2010, ApJ, 714, L280

\bibitem[\protect\citeauthoryear{{Gandhi} et~al.,}{{Gandhi}
  et~al.}{2013}]{gandhi_09js}
{Gandhi} P.  et~al., 2013, ApJ, 767, 166

\bibitem[\protect\citeauthoryear{{Hamuy}}{{Hamuy}}{2001}]{hamuy_thesis}
{Hamuy} M.,  2001, PhD thesis, The University of Arizona

\bibitem[\protect\citeauthoryear{Hamuy}{Hamuy}{2003}]{hamuy2003}
Hamuy M.,  2003, ApJ, 582, 905

\bibitem[\protect\citeauthoryear{{Hamuy}}{{Hamuy}}{2005}]{hamuy_scm2}
{Hamuy} M.,  2005, in {Marcaide} J.-M.,  {Weiler} K.~W.,  eds, IAU Colloq. 192:
  Cosmic Explosions, On the 10th Anniversary of SN1993J. p.~535

\bibitem[\protect\citeauthoryear{{Hamuy} \& {Pinto}}{{Hamuy} \&
  {Pinto}}{2002}]{hamuy_scm}
{Hamuy} M.,  {Pinto} P.~A.,  2002, ApJ, 566, L63

\bibitem[\protect\citeauthoryear{Hamuy et~al.,}{Hamuy et~al.}{2001}]{hamuy_epm}
Hamuy M.  et~al., 2001, ApJ, 558, 615

\bibitem[\protect\citeauthoryear{{Harutyunyan} et~al.,}{{Harutyunyan}
  et~al.}{2008}]{gelato}
{Harutyunyan} A.~H.  et~al., 2008, A\&A, 488, 383

\bibitem[\protect\citeauthoryear{{Hillier} \& {Dessart}}{{Hillier} \&
  {Dessart}}{2012}]{hillier_2012}
{Hillier} D.~J.,  {Dessart} L.,  2012, MNRAS, 424, 252

\bibitem[\protect\citeauthoryear{{Hillier} \& {Miller}}{{Hillier} \&
  {Miller}}{1998}]{hillier_cmfgen}
{Hillier} D.~J.,  {Miller} D.~L.,  1998, ApJ, 496, 407

\bibitem[\protect\citeauthoryear{Inserra et~al.,}{Inserra
  et~al.}{2011}]{inserra2011}
Inserra C.  et~al., 2011, MNRAS, 417, 261

\bibitem[\protect\citeauthoryear{Inserra et~al.,}{Inserra
  et~al.}{2012}]{inserra2012}
Inserra C.  et~al., 2012, MNRAS, 422, 1122

\bibitem[\protect\citeauthoryear{{Inserra} et~al.,}{{Inserra}
  et~al.}{2013}]{inserra2013}
{Inserra} C.  et~al., 2013, A\&A, 555, A142

\bibitem[\protect\citeauthoryear{Jones et~al.,}{Jones et~al.}{2009}]{jones2009}
Jones M.~I.  et~al., 2009, ApJ, 696, 1176

\bibitem[\protect\citeauthoryear{Kirshner \& Kwan}{Kirshner \&
  Kwan}{1974}]{epm_ref}
Kirshner R.~P.,  Kwan J.,  1974, ApJ, 193, 27

\bibitem[\protect\citeauthoryear{{Landolt}}{{Landolt}}{1992}]{landolt}
{Landolt} A.~U.,  1992, AJ, 104, 340

\bibitem[\protect\citeauthoryear{{Landolt} \& {Uomoto}}{{Landolt} \&
  {Uomoto}}{2007}]{landolt2007}
{Landolt} A.~U.,  {Uomoto} A.~K.,  2007, AJ, 133, 768

\bibitem[\protect\citeauthoryear{Leonard et~al.,}{Leonard
  et~al.}{2002a}]{leonard_99em}
Leonard D.~C.  et~al., 2002a, PASP, 114, 35

\bibitem[\protect\citeauthoryear{Leonard et~al.,}{Leonard
  et~al.}{2002b}]{leonard_99gi}
Leonard D.~C.  et~al., 2002b, AJ, 124, 2490

\bibitem[\protect\citeauthoryear{{Leonard}, {Kanbur}, {Ngeow} \&
  {Tanvir}}{{Leonard} et~al.}{2003}]{leonard_99em_ceph}
{Leonard} D.~C.,  {Kanbur} S.~M.,  {Ngeow} C.~C.,    {Tanvir} N.~R.,  2003,
  ApJ, 594, 247

\bibitem[\protect\citeauthoryear{{Li}, {Van Dyk}, {Filippenko} \&
  {Cuillandre}}{{Li} et~al.}{2005}]{li_04et}
{Li} W.,  {Van Dyk} S.~D.,  {Filippenko} A.~V.,    {Cuillandre} J.-C.,  2005,
  PASP, 117, 121

\bibitem[\protect\citeauthoryear{Li et~al.,}{Li et~al.}{2011}]{li}
Li W.  et~al., 2011, MNRAS, 412, 1441

\bibitem[\protect\citeauthoryear{Litvinova \& Nadyozhin}{Litvinova \&
  Nadyozhin}{1983}]{litvinova_models}
Litvinova I.~Y.,  Nadyozhin D.~K.,  1983, Astrophysics and Space Science, 89,
  89

\bibitem[\protect\citeauthoryear{{Livne}}{{Livne}}{1993}]{livne_v1d}
{Livne} E.,  1993, ApJ, 412, 634

\bibitem[\protect\citeauthoryear{{Maguire} et~al.,}{{Maguire}
  et~al.}{2010}]{maguire_04et}
{Maguire} K.  et~al., 2010, MNRAS, 404, 981

\bibitem[\protect\citeauthoryear{{Maguire}, {Kotak}, Smartt, Pastorello, Hamuy
  \& Bufano}{{Maguire} et~al.}{2010}]{maguire_scm}
{Maguire} K.,  {Kotak} R.,  Smartt S.~J.,  Pastorello A.,  Hamuy M.,    Bufano
  F.,  2010, MNRAS, 403, L11

\bibitem[\protect\citeauthoryear{{Maund} \& {Smartt}}{{Maund} \&
  {Smartt}}{2005}]{maund_progs}
{Maund} J.~R.,  {Smartt} S.~J.,  2005, MNRAS, 360, 288

\bibitem[\protect\citeauthoryear{{Maund}, {Mattila}, {Ramirez-Ruiz} \&
  {Eldridge}}{{Maund} et~al.}{2014}]{maund08bk}
{Maund} J.~R.,  {Mattila} S.,  {Ramirez-Ruiz} E.,    {Eldridge} J.~J.,  2014,
  MNRAS, 438, 1577

\bibitem[\protect\citeauthoryear{{Maund}, {Fraser}, {Reilly}, {Ergon} \&
  {Mattila}}{{Maund} et~al.}{2015}]{maund_progenitorsrevisited}
{Maund} J.~R.,  {Fraser} M.,  {Reilly} E.,  {Ergon} M.,    {Mattila} S.,  2015,
  MNRAS, 447, 3207

\bibitem[\protect\citeauthoryear{Milne et~al.,}{Milne et~al.}{2010}]{milne2010}
Milne, P.~A. et~al., 2010, ApJ, 721, 1627

\bibitem[\protect\citeauthoryear{Nugent et~al.,}{Nugent et~al.}{2006}]{nugent}
Nugent P.  et~al., 2006, ApJ, 645, 841

\bibitem[\protect\citeauthoryear{{Olivares} et~al.,}{{Olivares}
  et~al.}{2010}]{olivares_scm}
{Olivares} E.~F.  et~al., 2010, ApJ, 715, 833

\bibitem[\protect\citeauthoryear{{Pastorello} et~al.,}{{Pastorello}
  et~al.}{2004}]{pastorello99br}
{Pastorello} A.  et~al., 2004, MNRAS, 347, 74

\bibitem[\protect\citeauthoryear{Pastorello et~al.,}{Pastorello
  et~al.}{2009}]{pastorello05csII}
Pastorello A.  et~al., 2009, MNRAS, 394, 2266

\bibitem[\protect\citeauthoryear{{Pastorello} et~al.,}{{Pastorello}
  et~al.}{2012}]{pastorello_09E}
{Pastorello} A.  et~al., 2012, A\&A, 537, A141

\bibitem[\protect\citeauthoryear{{Patat}, {Barbon}, {Cappellaro} \&
  {Turatto}}{{Patat} et~al.}{1994}]{patat_1994}
{Patat} F.,  {Barbon} R.,  {Cappellaro} E.,    {Turatto} M.,  1994, A\&A, 282,
  731

\bibitem[\protect\citeauthoryear{{Paxton}, {Bildsten}, {Dotter}, {Herwig},
  {Lesaffre} \& {Timmes}}{{Paxton} et~al.}{2011}]{paxton_mesa}
{Paxton} B.,  {Bildsten} L.,  {Dotter} A.,  {Herwig} F.,  {Lesaffre} P.,
  {Timmes} F.,  2011, ApJS, 192, 3

\bibitem[\protect\citeauthoryear{{Paxton} et~al.,}{{Paxton}
  et~al.}{2013}]{paxton_mesa2}
{Paxton} B.  et~al., 2013, ApJS, 208, 4

\bibitem[\protect\citeauthoryear{{Pettini} \& {Pagel}}{{Pettini} \&
  {Pagel}}{2004}]{pettini_metallicity}
{Pettini} M.,  {Pagel} B.~E.~J.,  2004, MNRAS, 348, L59

\bibitem[\protect\citeauthoryear{{Phillips} et~al.,}{{Phillips}
  et~al.}{2013}]{phillips_extinction}
{Phillips} M.~M.  et~al., 2013, ApJ, 779, 38

\bibitem[\protect\citeauthoryear{{Pignata} et~al.,}{{Pignata}
  et~al.}{2009}]{2009ib_felfed}
{Pignata} G.  et~al., 2009, Central Bureau Electronic Telegrams, 1902, 1

\bibitem[\protect\citeauthoryear{Poznanski et~al.,}{Poznanski
  et~al.}{2009}]{poznanski_scm}
Poznanski D.  et~al., 2009, ApJ, 694, 1067

\bibitem[\protect\citeauthoryear{Poznanski, Ganeshalingam, Silverman \&
  Filippenko}{Poznanski et~al.}{2011}]{poznanski2011}
Poznanski D.,  Ganeshalingam M.,  Silverman J.~M.,    Filippenko A.~V.,  2011,
  MNRAS, 415, L81

\bibitem[\protect\citeauthoryear{Poznanski, Prochaska \& Bloom}{Poznanski
  et~al.}{2012}]{poznanski2012}
Poznanski D.,  Prochaska J.~X.,    Bloom J.~S.,  2012, MNRAS, 426, 1465

\bibitem[\protect\citeauthoryear{{Pumo} \& {Zampieri}}{{Pumo} \&
  {Zampieri}}{2011}]{pumo2011}
{Pumo} M.~L.,  {Zampieri} L.,  2011, ApJ, 741, 41

\bibitem[\protect\citeauthoryear{{Pumo} \& {Zampieri}}{{Pumo} \&
  {Zampieri}}{2013}]{pumo2013}
{Pumo} M.~L.,  {Zampieri} L.,  2013, MNRAS, 434, 3445

\bibitem[\protect\citeauthoryear{{Pumo}, {Zampieri} \& {Turatto}}{{Pumo}
  et~al.}{2010}]{pumo2010}
{Pumo} M.~L.,  {Zampieri} L.,    {Turatto} M.,  2010, MSAIS, 14, 123

\bibitem[\protect\citeauthoryear{Roy et~al.,}{Roy et~al.}{2011}]{roy08in}
Roy R.  et~al., 2011, ApJ, 736, 76

\bibitem[\protect\citeauthoryear{{Sanders} et~al.,}{{Sanders}
  et~al.}{2015}]{sanders_2014}
{Sanders} N.~E.  et~al., 2015, ApJ, 799, 208

\bibitem[\protect\citeauthoryear{{Schlafly} \& {Finkbeiner}}{{Schlafly} \&
  {Finkbeiner}}{2011}]{reddening_recalib}
{Schlafly} E.~F.,  {Finkbeiner} D.~P.,  2011, ApJ, 737, 103

\bibitem[\protect\citeauthoryear{Schmidt et~al.,}{Schmidt
  et~al.}{1994}]{Schmidt1994}
Schmidt B.~P.  et~al., 1994, The Astronomical Journal, 107, 1444

\bibitem[\protect\citeauthoryear{{Skrutskie} et~al.,}{{Skrutskie}
  et~al.}{2006}]{2mass_catalogue}
{Skrutskie} M.~F.  et~al., 2006, AJ, 131, 1163

\bibitem[\protect\citeauthoryear{Smartt}{Smartt}{2009}]{smartt_rev}
Smartt S.~J.,  2009, ARA\&A, 47, 63

\bibitem[\protect\citeauthoryear{Smartt, Eldridge, Crockett \& Maund}{Smartt
  et~al.}{2009}]{smartt2009}
Smartt S.~J.,  Eldridge J.~J.,  Crockett R.~M.,    Maund J.~R.,  2009, MNRAS,
  395, 1409

\bibitem[\protect\citeauthoryear{{Smith} et~al.,}{{Smith}
  et~al.}{2002}]{sloan_std_fields}
{Smith} J.~A.  et~al., 2002, AJ, 123, 2121

\bibitem[\protect\citeauthoryear{{Spiro} et~al.,}{{Spiro}
  et~al.}{2014}]{spiro_2014}
{Spiro} S.  et~al., 2014, MNRAS, 439, 2873

\bibitem[\protect\citeauthoryear{{Springob}, {Masters}, {Haynes}, {Giovanelli}
  \& {Marinoni}}{{Springob} et~al.}{2007}]{TF3_ngc1559}
{Springob} C.~M.,  {Masters} K.~L.,  {Haynes} M.~P.,  {Giovanelli} R.,
  {Marinoni} C.,  2007, ApJS, 172, 599

\bibitem[\protect\citeauthoryear{{Stritzinger} \& {Folatelli}}{{Stritzinger} \&
  {Folatelli}}{2009}]{2009ib_classification}
{Stritzinger} M.,  {Folatelli} G.,  2009, Central Bureau Electronic Telegrams,
  1902, 2

\bibitem[\protect\citeauthoryear{Tak\'{a}ts \& Vink\'{o}}{Tak\'{a}ts \&
  Vink\'{o}}{2006}]{takats2006}
Tak\'{a}ts K.,  Vink\'{o} J.,  2006, MNRAS, 372, 1735

\bibitem[\protect\citeauthoryear{Tak\'{a}ts \& Vink\'{o}}{Tak\'{a}ts \&
  Vink\'{o}}{2012}]{takats2012}
Tak\'{a}ts K.,  Vink\'{o} J.,  2012, MNRAS, 419, 2783

\bibitem[\protect\citeauthoryear{{Tak{\'a}ts} et~al.,}{{Tak{\'a}ts}
  et~al.}{2014}]{takats2014}
{Tak{\'a}ts} K.  et~al., 2014, MNRAS, 438, 368

\bibitem[\protect\citeauthoryear{{Tomasella} et~al.,}{{Tomasella}
  et~al.}{2013}]{tomasella_12A}
{Tomasella} L.  et~al., 2013, MNRAS, 434, 1636

\bibitem[\protect\citeauthoryear{{Tully} \& {Courtois}}{{Tully} \&
  {Courtois}}{2012}]{TF2_ngc1559}
{Tully} R.~B.,  {Courtois} H.~M.,  2012, ApJ, 749, 78

\bibitem[\protect\citeauthoryear{{Tully} \& {Fisher}}{{Tully} \&
  {Fisher}}{1977}]{tullyfisher}
{Tully} R.~B.,  {Fisher} J.~R.,  1977, A\&A, 54, 661

\bibitem[\protect\citeauthoryear{Turatto et~al.,}{Turatto
  et~al.}{1998}]{turatto_ba}
Turatto M.  et~al., 1998, ApJ, 498, L129

\bibitem[\protect\citeauthoryear{Utrobin}{Utrobin}{2007}]{utrobin99em}
Utrobin V.~P.,  2007, A\&A, 461, 233

\bibitem[\protect\citeauthoryear{Utrobin, Chugai \& Pastorello}{Utrobin
  et~al.}{2007}]{utrobin03Z}
Utrobin V.~P.,  Chugai N.~N.,    Pastorello A.,  2007, A\&A, 475, 973

\bibitem[\protect\citeauthoryear{{Van Dyk} et~al.,}{{Van Dyk}
  et~al.}{2012}]{vandyk12aw}
{Van Dyk} S.~D.  et~al., 2012, ApJ, 756, 131

\bibitem[\protect\citeauthoryear{Vink\'{o} et~al.,}{Vink\'{o}
  et~al.}{2012}]{vinko11dh}
Vink\'{o} J.  et~al., 2012, A\&A, 540, A93

\bibitem[\protect\citeauthoryear{{Walmswell} \& {Eldridge}}{{Walmswell} \&
  {Eldridge}}{2012}]{walmswell_eldridge}
{Walmswell} J.~J.,  {Eldridge} J.~J.,  2012, MNRAS, 419, 2054

\bibitem[\protect\citeauthoryear{{Wang} \& {Baade}}{{Wang} \&
  {Baade}}{2005}]{2005df_discovery}
{Wang} L.,  {Baade} D.,  2005, Central Bureau Electronic Telegrams, 193, 1

\bibitem[\protect\citeauthoryear{{Willick}, {Courteau}, {Faber}, {Burstein},
  {Dekel} \& {Strauss}}{{Willick} et~al.}{1997}]{TF1_ngc1559}
{Willick} J.~A.,  {Courteau} S.,  {Faber} S.~M.,  {Burstein} D.,  {Dekel} A.,
   {Strauss} M.~A.,  1997, ApJS, 109, 333

\bibitem[\protect\citeauthoryear{{Zampieri}}{{Zampieri}}{2007}]{zampieri_sne_conf}
{Zampieri} L.,  2007, in {di Salvo} T.,  {Israel} G.~L.,  {Piersant} L.,
  {Burderi} L.,  {Matt} G.,  {Tornambe} A.,   {Menna} M.~T.,  eds,  American
  Institute of Physics Conference Series Vol. 924, The Multicolored Landscape
  of Compact Objects and Their Explosive Origins. pp 358--365

\bibitem[\protect\citeauthoryear{{Zampieri}, {Pastorello}, {Turatto},
  {Cappellaro}, {Benetti}, {Altavilla}, {Mazzali} \& {Hamuy}}{{Zampieri}
  et~al.}{2003}]{zampieri2003}
{Zampieri} L.,  {Pastorello} A.,  {Turatto} M.,  {Cappellaro} E.,  {Benetti}
  S.,  {Altavilla} G.,  {Mazzali} P.,    {Hamuy} M.,  2003, MNRAS, 338, 711

\end{thebibliography}

\appendix

\section{Photometric tables of SN 2009ib}\label{app_a}

In this Section we include the the magnitudes of the local sequence stars that we used to calibrate the photometric measurements of SN~2009ib (Tables~\ref{seq_bvri}, \ref{seq_ugriz}, \ref{seq_nir}), and the calibrated light curves of the SN (Tables~\ref{lc_landolt}, \ref{lc_sloan}, \ref{lc_nir}). All these tables are available in a machine-readable format in the online version of the paper.

 \begin{table*}
 \centering
 \begin{minipage}{\textwidth}
  \caption{$UBVRI$ magnitudes of the local sequence of stars used for the calibration.}
  \label{seq_bvri}
  \begin{tabular}{@{}cccccccc@{}}
  \hline
  \hline
Star & $\alpha_{2000}$ & $\delta_{2000}$ & $U$  & $B$ & $V$ & $R$ & $I$ \\
  \hline
 1 & 04:17:07.209 &  -62:46:13.44 & &  19.367     (0.078) &  17.850     (0.038)  &  16.830     (0.031)  &  15.690     (0.024) \\ 
 2 & 04:17:05.345 &  -62:45:39.15 & &  16.663     (0.038) &  15.754     (0.030)  &  15.212     (0.022)  &  14.757     (0.011) \\
 3 & 04:17:28.126 &  -62:45:47.45 & &  18.610     (0.042) &  17.569     (0.027)  &  16.997     (0.044)  &  16.526     (0.023) \\
 4 & 04:17:38.715 &  -62:45:32.07 & &  18.608     (0.040) &  17.116     (0.044)  &  16.203     (0.045)  &  15.359     (0.016) \\
 5 & 04:17:04.664 &  -62:45:52.59 & &  17.047     (0.042) &  16.437     (0.027)  &  16.094     (0.018)  &  15.733     (0.011) \\ 
 6 & 04:18:02.142 &  -62:45:30.44 & &  17.632     (0.034) &  16.641     (0.038)  &  16.066     (0.024)  &  15.542     (0.014) \\ 
 7 & 04:18:01.888 &  -62:45:58.22 & &  15.674     (0.017) &  14.989     (0.026)  &  14.601     (0.024)  &  14.234     (0.018) \\ 
 8 & 04:18:06.207 &  -62:45:57.24 & &  18.252     (0.049) &  17.759     (0.033)  &  17.489     (0.042)  &  17.151     (0.050) \\ 
 9 & 04:18:02.508 &  -62:47:20.28 & &  17.508     (0.032) &  16.218     (0.028)  &  15.412     (0.034)  &  14.758     (0.010) \\ 
10 & 04:18:11.929 &  -62:48:49.89 & &  16.690     (0.025) &  15.569     (0.034)  &  14.831     (0.033)  &  14.258     (0.022) \\ 
11 & 04:18:06.386 &  -62:49:30.78 & &  19.033     (0.056) &  18.045     (0.029)  &  17.354     (0.029)  &  16.811     (0.027) \\ 
12 & 04:17:51.943 &  -62:50:34.43 & &  13.703     (0.017) &  13.038     (0.019)  &  12.671     (0.025)  &  12.332     (0.014) \\ 
13 & 04:17:40.451 &  -62:50:46.79 & &  13.818     (0.020) &  13.300     (0.018)  &  12.996     (0.014)  &  12.705     (0.012) \\ 
14 & 04:17:29.532 &  -62:50:08.59 & &  15.561     (0.014) &  14.948     (0.018)  &  14.590     (0.022)  &  14.249     (0.016) \\ 
15 & 04:17:32.022 &  -62:49:05.93 & &  18.418     (0.037) &  17.592     (0.022)  &  17.100     (0.026)  &  16.679     (0.023) \\ 
16 & 04:17:06.264 &  -62:48:48.01 & &  16.909     (0.013) &  16.161     (0.033)  &  15.692     (0.020)  &  15.266     (0.014) \\ 
17 & 04:17:07.593 &  -62:47:35.37 & &  18.212     (0.028) &  17.386     (0.037)  &  16.891     (0.035)  &  16.434     (0.018) \\ 
18 & 04:17:05.610 &  -62:47:15.30 & &  18.809     (0.023) &  18.216     (0.060)  &  17.819     (0.020)  &  17.471     (0.037) \\ 
19 & 04:17:53.330 & -62:46:48.26 & 16.524 (0.101) & 15.666 (0.015) & 14.713 (0.008) & 14.462 (0.238) & 13.753 (0.026)\\
20 & 04:17:44.525 & -62:45:42.38 & 19.372 (0.192) & 19.199 (0.085) & 18.466 (0.054) & 18.127 (0.019) & 17.655 (0.040) \\
21 & 04:17:46.119 & -62:47:44.06 & 20.591 (0.459) & 19.173 (0.012) & 18.105 (0.016) & 17.502 (0.018) & 16.860 (0.020) \\
22 & 04:17:40.856 & -62:48:20.15 & 20.120 (0.047) & 19.709 (0.050) &  18.977 (0.054) & 18.579 (0.052) & 18.087 (0.081) \\
\hline				  
\end{tabular}			  
\end{minipage}
\end{table*}

 \begin{table*}
 \centering
 \begin{minipage}{\textwidth}
  \caption{$u'g'r'i'z'$ magnitudes of the local sequence of stars used for the calibration.}
  \label{seq_ugriz}
  \begin{tabular}{@{}cccccccc@{}}
  \hline
  \hline
Star & $\alpha_{2000}$ & $\delta_{2000}$  & u' & g' & r' & i' & z' \\
  \hline
 1 &  04:17:07.209 &  -62:46:13.44 &                       &   18.525     (0.099)  &   17.284     (0.034)  &   16.310     (0.020)  &   15.865     (0.064)  \\
 2 &  04:17:05.345 &  -62:45:39.15 &                       &   16.216     (0.020)  &   15.468     (0.023)  &   15.215     (0.021)  &   15.097     (0.035)  \\
 3 &  04:17:28.126 &  -62:45:47.45 &                       &   18.104     (0.047)  &   17.296     (0.032)  &   17.002     (0.026)  &   16.878     (0.046)  \\
 4 &  04:17:38.715 &  -62:45:32.07 &                       &   17.831     (0.072)  &   16.604     (0.034)  &   15.942     (0.033)  &   15.611     (0.059)  \\
 5 &  04:17:04.664 &  -62:45:52.59 &                       &   16.729     (0.021)  &   16.308     (0.023)  &   16.163     (0.018)  &   16.129     (0.027)  \\
 6 &  04:18:02.142 &  -62:45:30.44 &                       &   17.142     (0.021)  &   16.338     (0.034)  &   16.019     (0.032)  &   15.868     (0.046)  \\
 7 &  04:18:01.888 &  -62:45:58.22 &                       &   16.673     (0.057)  &   15.310     (0.019)  &   14.843     (0.026)  &   14.678     (0.026)  \\ 
 8 &  04:18:06.207 &  -62:45:57.24 &                       &   18.008     (0.035)  &   17.701     (0.028)  &   17.574     (0.016)  &   17.590     (0.052)  \\
 9 &  04:18:02.508 &  -62:47:20.28 &                       &   16.859     (0.036)  &   15.743     (0.031)  &   15.279     (0.034)  &   15.050     (0.046)  \\
10 &  04:18:11.929 &  -62:48:49.89 &                       &   16.145     (0.019)  &   15.140     (0.031)  &   14.740     (0.027)  &   14.557     (0.036)  \\
11 &  04:18:06.386 &  -62:49:30.78 &                       &   18.561     (0.042)  &   17.648     (0.038)  &   17.282     (0.034)  &   17.133     (0.041)  \\
12 &  04:17:51.943 &  -62:50:34.43 &    14.689     (0.043)  &   13.326     (0.031)  &   12.906     (0.016)  &   12.761     (0.020)  &   12.730     (0.019)  \\
13 &  04:17:40.451 &  -62:50:46.79 &    14.699     (0.052)  &   13.528     (0.033)  &   13.216     (0.007)  &   13.124     (0.017)  &   13.122     (0.021)  \\
14 &  04:17:29.532 &  -62:50:08.59 &    16.514     (0.039)  &   15.235     (0.019)  &   14.822     (0.004)  &   14.683     (0.018)  &   14.645     (0.020)  \\
15 &  04:17:32.022 &  -62:49:05.93 &                       &   17.958     (0.062)  &   17.376     (0.017)  &   17.134     (0.020)  &   17.065     (0.043)  \\
16 &  04:17:06.264 &  -62:48:48.01 &                       &   18.097     (0.074)  &   16.516     (0.040)  &   15.947     (0.017)  &   15.719     (0.016)  \\
17 &  04:17:07.593 &  -62:47:35.37 &                       &   17.825     (0.056)  &   17.143     (0.024)  &   16.902     (0.010)  &   16.797     (0.033)  \\
18 &  04:17:05.610 &  -62:47:15.30 &                       &   18.508     (0.052)  &   18.070     (0.032)  &   17.925     (0.037)  &   17.905     (0.054)  \\

\hline				  
\end{tabular}			  
\end{minipage}
\end{table*}

 \begin{table*}
 \centering
 \begin{minipage}{\textwidth}
  \caption{Magnitudes of the local sequence of stars used for the calibration of NIR photometry taken from the 2MASS catalogue \citep{2mass_catalogue}.}
  \label{seq_nir}
  \begin{tabular}{@{}ccccc@{}}
  \hline
  \hline
Star & $\alpha_{2000}$ & $\delta_{2000}$  & $J$ & $H$  \\
  \hline
A & 04:17:36.46 & -62:42:41.79 & 11.723 (0.027) & 11.216 (0.024)  \\
B & 04:17:43.44 & -62:43:57.30 & 12.409 (0.027) & 12.051 (0.024)  \\
C & 04:18:01.92 & -62:45:58.35 & 13.703 (0.030) & 13.331 (0.030)  \\
D & 04:17:53.33 & -62:46:48.26 & 13.068 (0.026) & 12.572 (0.026)  \\
E & 04:18:02.56 & -62:47:20.28 & 13.945 (0.032) & 13.219 (0.030)  \\
F & 04:18:11.95 & -62:48:50.11 & 13.399 (0.027) & 12.768 (0.026)  \\
G & 04:17:51.92 & -62:50:34.94 & 11.826 (0.026) & 11.500 (0.024)  \\
H & 04:17:40.48 & -62:50:46.96 & 12.303 (0.028) & 12.010 (0.024)  \\
I & 04:17:29.58 & -62:50:08.81 & 13.790 (0.027) & 13.417 (0.030)  \\
J & 04:17:19.90 & -62:47:51.73 & 11.848 (0.027) & 11.570 (0.023)  \\
\hline				  
\end{tabular}			  
\end{minipage}
\end{table*}

 \begin{table*}
 \centering
 \begin{minipage}{\textwidth}
  \caption{$UBVRI$ magnitudes of SN 2009ib.}
  \label{lc_landolt}
  \begin{tabular}{@{}llcccccc@{}}
  \hline
  \hline
Date & MJD & $U$ & $B$ & $V$ & $R$ & $I$ & Telescope \\
  \hline
2009/08/11  & 55054.4 &                 &  16.030  (0.016) & 15.854 (0.015)  &  15.612 (0.014)  &  15.499 (0.014) & PROMPT \\
2009/08/13  & 55057.3 &                 &  16.241  (0.157) & 15.810 (0.263)  &  15.617 (0.271)  &  15.491 (0.181) & NTT \\
2009/08/20  & 55063.3 &                 &  16.431  (0.042) & 15.875 (0.031)  &  15.547 (0.033)  &  15.409 (0.059) & CTIO  \\
2009/08/21  & 55064.4 &                 &  16.569  (0.020) & 15.920 (0.015)  &  15.575 (0.015)  &  15.439 (0.013) & PROMPT \\
2009/08/23  & 55066.3&                  &  16.657  (0.053) & 15.931 (0.065)  &  15.562 (0.049)  &  15.381 (0.061) & CTIO \\ 
2009/08/23  & 55066.4 &                 &  16.694  (0.020) & 15.964 (0.015)  &  15.599 (0.013)  &  15.420 (0.013) & CTIO\\
2009/08/24  & 55067.4 &                 &  16.748  (0.027) & 15.975 (0.015)  &  15.584 (0.013)  &  15.410 (0.013) & PROMPT \\ 
2009/08/25  & 55068.4 &                 &  16.780  (0.025) & 15.986 (0.014)  &  15.632 (0.015)  &  15.418 (0.015) & PROMPT \\
2009/08/26  & 55069.3 &                 &  16.771  (0.023) & 15.942 (0.042)  &  15.555 (0.047)  &  15.373 (0.056) & CTIO\\
2009/08/27  & 55070.4 &                 &  16.863  (0.029) & 16.019 (0.015)  &  15.649 (0.015)  &  15.415 (0.013) & PROMPT \\ 
2009/08/29  & 55072.3 &                 &  16.957  (0.026) & 16.041 (0.017)  &  15.675 (0.014)  &  15.430 (0.015) & PROMPT \\
2009/08/29  & 55072.4 &                 &  16.891  (0.043) & 15.975 (0.029)  &  15.592 (0.047)  &  15.377 (0.024) & CTIO\\
2009/08/31  & 55074.3 &                 &  17.014  (0.020) & 16.076 (0.016)  &  15.649 (0.013)  &  15.426 (0.013 & PROMPT \\
2009/09/01  & 55075.4 &                 &  16.985  (0.047) & 16.036 (0.030)  &  15.596 (0.022)  &  15.371 (0.015)  & CTIO\\
2009/09/03  & 55077.3 &                 &  17.090  (0.034) & 16.110 (0.018)  &  15.685 (0.015)  &  15.401 (0.022) & PROMPT \\ 
2009/09/05  & 55079.4 &                 &  17.146  (0.035) & 16.123 (0.025)  &  15.664 (0.013)  &  15.403 (0.038) & PROMPT \\
2009/09/08  & 55082.3 &                 &  17.147  (0.099) & 16.116 (0.054)  &  15.636 (0.051)  &  15.377 (0.039) & CTIO\\
2009/09/09  & 55083.3 &                 &  17.218  (0.023) & 16.121 (0.014)  &  15.696 (0.013)  &  15.396 (0.013) & PROMPT \\ 
2009/09/10  & 55084.3 &                 &  17.215  (0.031) &                &                 &                & PROMPT \\ 
2009/09/11  & 55085.3 &                 &  17.259  (0.034) & 16.122 (0.015)  &  15.684 (0.014)  &  15.391 (0.013) & PROMPT \\
2009/09/11  & 55085.4 &                 &  17.286  (0.041) &  16.134 (0.024) &  15.639 (0.025)  &  15.324 (0.015) & CTIO\\
2009/09/12  & 55086.3 &                 &  17.293  (0.018) & 16.121 (0.014)  &  15.675 (0.013)  &  15.373 (0.013) & PROMPT \\ 
2009/09/14  & 55088.4 &                 &  17.301  (0.024) & 16.126 (0.025)  &  15.690 (0.019)  &  15.356 (0.031) & PROMPT \\
2009/09/14  & 55088.4 &                 &  17.248  (0.042) & 16.086 (0.029)  &  15.632 (0.033)  &  15.362 (0.022) & CTIO\\
2009/09/16  & 55090.3 &                 &  17.316  (0.024) & 16.157 (0.016)  &  15.720 (0.015)  &  15.372 (0.013) & PROMPT \\ 
2009/09/18  & 55092.3 &                 &  17.330  (0.045) & 16.178 (0.033)  &  15.758 (0.071)  &  15.364 (0.102) & PROMPT \\ 
2009/09/20  & 55094.3 &                 &                 & 16.184 (0.016)  &  15.726 (0.015)  &  15.388 (0.014) & PROMPT \\ 
2009/09/22  & 55096.3 &                 &  17.402  (0.022) & 16.221 (0.015)  &  15.727 (0.015)  &  15.398 (0.013) & PROMPT \\
2009/09/22  & 55096.3 &                 &  17.389  (0.046) & 16.179 (0.053)  &  15.658 (0.028)  &  15.385 (0.033) & CTIO \\
2009/09/24  & 55098.3 &                 &  17.423  (0.034) & 16.219 (0.014)  &  15.720 (0.013)  &  15.396 (0.012) & PROMPT \\ 
2009/09/26  & 55100.3 &                 &  17.438  (0.031) & 16.216 (0.018)  &  15.739 (0.016)  &  15.384 (0.015) & PROMPT \\
2009/09/30  & 55104.3 &                 &  17.499  (0.061) & 16.195 (0.024)  &  15.708 (0.021)  &  15.332 (0.020)  & CTIO\\
2009/10/05  & 55109.3 &                 &  17.516  (0.022) &                &                 &                & PROMPT \\ 
2009/10/07  & 55111.3 &                 &  17.545  (0.021) &                &                 &                & PROMPT \\ 
2009/10/09  & 55113.3 &                 &  17.506  (0.027) & 16.241 (0.015)  &  15.730 (0.013)  &  15.347 (0.013) & PROMPT \\ 
2009/10/12  & 55116.3 &                 &  17.523  (0.038) & 16.216 (0.016)  &  15.728 (0.013)  &  15.362 (0.012) & PROMPT \\ 
2009/10/13  & 55117.3 &                 &  17.528  (0.109) &                &                 &                & PROMPT \\ 
2009/10/17  & 55121.3 &                 &                 & 16.204 (0.014)  &  15.703 (0.013)  &  15.344 (0.012) & PROMPT \\ 
2009/10/19  & 55124.4 & 19.425  (0.349) &  17.531  (0.157) & 16.213 (0.145)  &  15.715 (0.243)  &  15.323 (0.106) & NTT \\ 
2009/10/27  & 55131.3 &                 &  17.506  (0.040) & 16.206 (0.020)  &  15.683 (0.017)  &  15.328 (0.016) & PROMPT \\ 
2009/10/30  & 55134.3 &                 &                 & 16.204 (0.026)  &                 &                & PROMPT \\ 
2009/11/03  & 55138.2 &                 &  17.601  (0.020) & 16.255 (0.015)  &  15.731 (0.013)  &  15.344 (0.012) & PROMPT \\ 
2009/11/07  & 55142.3 &                 &  17.576  (0.030) &                &                 &                & PROMPT \\ 
2009/11/08  & 55143.3 &                 &                 & 16.192 (0.018)  &  15.711 (0.015)  &  15.359 (0.013) & PROMPT \\ 
2009/11/12  & 55147.2 &                 &  17.583  (0.039) &                &                 &                & PROMPT \\ 
2009/11/13  & 55148.2 &                 &                 & 16.262 (0.017)  &  15.729 (0.015)  &  15.384 (0.015) & PROMPT \\ 
2009/11/16  & 55151.1 &                 &  17.645  (0.029) & 16.302 (0.014)  &  15.758 (0.013)  &  15.399 (0.012) & PROMPT \\ 
2009/11/22  & 55157.2 & 19.741  (0.455)  &  17.749  (0.188) & 16.369 (0.184)  &  15.874 (0.201)  &  15.457 (0.179) & NTT \\ 
2009/11/23  & 55158.1 &                 &                 &                &  15.812 (0.019)  &  15.448 (0.018) & PROMPT \\ 
2009/11/24  & 55159.2 &                 &                 & 16.365 (0.020)  &  15.806 (0.013)  &  15.458 (0.012) & PROMPT \\ 
2009/12/02  & 55167.1 &                 &                 & 16.478 (0.017)  &                 &                & PROMPT \\ 
2009/12/03  & 55168.1 &                 &  17.891  (0.039) &                &                 &                & PROMPT \\ 
2009/12/07  & 55172.1 &                 &  17.950  (0.062) & 16.602 (0.018)  &  15.983 (0.017)  &  15.632 (0.014) & PROMPT \\ 
2009/12/09  & 55174.1 &                 &  18.054  (0.050) &                &                 &                & PROMPT \\ 
2009/12/10  & 55175.2 &                 &  18.039  (0.080) & 16.697 (0.018)  &  16.071 (0.013)  &  15.721 (0.013) & PROMPT \\
\hline				                            
\end{tabular}			  
\end{minipage}
\end{table*}

\begin{table*}
 \centering
 \begin{minipage}{170mm}
  \contcaption{}
  \label{}
  \begin{tabular}{@{}cccccccc@{}}
  \hline
  \hline
 Date & MJD & $U$ & $B$ & $V$ & $R$ & $I$ & Telescope \\
\hline
2009/12/12  & 55177.1 &                 &  18.152  (0.064) &                &                 &                & PROMPT \\ 
2009/12/14  & 55179.2 &                 &  18.376  (0.061) &                &                 &                & PROMPT \\ 
2009/12/15  & 55180.1 &                 &  18.393  (0.059) &                &                 &                & PROMPT \\ 
2009/12/16  & 55181.1 &                 &                 & 17.098 (0.024)  &  16.394 (0.017)  &  15.999 (0.015) & PROMPT \\ 
2009/12/18  & 55183.0 &                 &  18.692  (0.065) & 17.298 (0.030)  &                 &                & PROMPT \\ 
2009/12/22  & 55187.3 &                 &                 & 17.570 (0.022)  &  16.809 (0.021)  &  16.402 (0.015) & PROMPT \\ 
2009/12/24  & 55189.0 &                 &                 & 17.680 (0.023)  &  16.864 (0.027)  &                & PROMPT \\ 
2009/12/24  & 55190.2 &                 &  19.266  (0.260) & 17.768 (0.137)  &  16.986 (0.250)  &  16.559 (0.159) & NTT \\
2010/01/04  & 55200.1 &                 &  19.206  (0.109) &                &                 &                & PROMPT \\ 
2010/01/07  & 55203.0 &                 &                 & 17.931 (0.090)  &                 &                & PROMPT \\ 
2010/01/08  & 55204.0 &                 &                 &                &  17.035 (0.036)  &                & PROMPT \\
2010/01/12  & 55208.1 &                 &  19.250  (0.168) &                &                 &                & PROMPT \\
2010/01/17  & 55213.1 &                 &  19.282  (0.121) &                &                 &                & PROMPT \\
2010/01/18  & 55214.2 &                 &                 & 18.003 (0.032)  &                 &                 & PROMPT\\
2010/01/24  & 55221.2 &                 &                 & 17.999 (0.190)  &  17.140 (0.284)  &  16.636 (0.199) & NTT \\
2010/04/17  & 55304.0 &                 &                 & 18.702 (0.221)  &  18.143 (0.373)  &  18.207 (0.231) & NTT \\
\hline				                            
\end{tabular}			  
 \begin{tablenotes}
       \item[a]{Telescopes: PROMPT - Panchromatic Robotic Optical Monitoring and Polarimetry Telescopes; CTIO - CTIO 1.3 m SMARTS telescope + ANDICAM; NTT - 3.6-m New Technology Telescope + EFOSC2.}
     \end{tablenotes}
\end{minipage}
\end{table*}

\begin{table*}
 \centering
 \begin{minipage}{\textwidth}
  \caption{$u'g'r'i'z'$ magnitudes of SN 2009ib obtained with the PROMPT telescopes.}
  \label{lc_sloan}
  \begin{tabular}{@{}ccccccc@{}}
  \hline
  \hline
Date & MJD & $u'$ & $g'$ & $r'$ & $i'$ & $z'$  \\
  \hline
2009/08/11 &   55054.4 &   16.148 (0.035)&   15.921 (0.015)&   15.781 (0.013)&   15.869    (0.015)&   15.935 (0.019) \\
2009/08/21 &   55064.4 &   17.651 (0.069)&   16.222 (0.015)&   15.797 (0.013)&   15.840    (0.015)&   15.836 (0.018) \\
2009/08/23 &   55066.4 &   17.791 (0.084)&   16.293 (0.015)&   15.828 (0.012)&   15.840    (0.014)&   15.820 (0.017) \\
2009/08/24 &   55067.4 &   18.022 (0.074)&   16.314 (0.015)&   15.830 (0.012)&   15.838    (0.026)&                 \\
2009/08/25 &   55068.4 &                &   16.352 (0.017)&                &                   &                 \\
2009/08/26 &   55069.4 &   18.204 (0.087)&                &                &                   &                 \\
2009/08/27 &   55070.4 &                &   16.420 (0.019)&                &                   &                 \\
2009/08/28 &   55071.3 &   18.306 (0.080)&                &                &                   &                 \\
2009/08/29 &   55072.3 &                &   16.458 (0.016)&                &                   &                 \\
2009/08/30 &   55073.4 &   18.517 (0.068)&                &                &                   &                 \\
2009/08/31 &   55074.3 &                &   16.486 (0.015)&                &                   &                 \\
2009/09/01 &   55075.3 &   18.600 (0.088)&                &                &                   &                 \\
2009/09/05 &   55079.4 &                &   16.548 (0.019)&   15.910 (0.015)&   15.850    (0.018)&   15.788 (0.053) \\
2009/09/09 &   55083.4 &                &   16.582 (0.014)&   15.912 (0.013)&   15.854    (0.013)&   15.765 (0.015) \\
2009/09/11 &   55085.3 &                &   16.594 (0.014)&   15.913 (0.013)&   15.858    (0.014)&   15.770 (0.015) \\
2009/09/13 &   55087.4 &                &   16.635 (0.014)&   15.910 (0.013)&   15.847    (0.015)&   15.766 (0.059) \\
2009/09/14 &   55088.4 &                &                &                &                   &   15.757 (0.026) \\
2009/09/15 &   55089.3 &                &   16.654 (0.016)&                &                   &                 \\
2009/09/16 &   55090.4 &                &                &                &                   &   15.750 (0.021) \\
2009/09/17 &   55091.3 &                &   16.663 (0.014)&   15.924 (0.013)&   15.864    (0.014)&                 \\
2009/09/18 &   55092.4 &                &                &                &   15.831    (0.026)&                 \\
2009/09/19 &   55093.2 &                &   16.693 (0.015)&   15.921 (0.012)&   15.854    (0.012)&   15.747 (0.014) \\
2009/09/21 &   55095.3 &                &   16.720 (0.017)&   15.940 (0.016)&   15.853    (0.021)&   15.752 (0.018) \\
2009/09/23 &   55097.4 &                &   16.721 (0.017)&   15.943 (0.013)&   15.852    (0.014)&   15.753 (0.015) \\
2009/09/25 &   55099.3 &                &   16.735 (0.016)&   15.953 (0.013)&   15.866    (0.013)&   15.745 (0.015) \\
2009/10/01 &   55105.3 &                &   16.766 (0.014)&   15.955 (0.016)&   15.867    (0.018)&   15.740 (0.029) \\
2009/10/05 &   55109.3 &                &   16.774 (0.016)&                &                   &                 \\
2009/10/07 &   55111.3 &                &   16.784 (0.015)&                &                   &                 \\
2009/10/08 &   55112.3 &                &   16.783 (0.016)&   15.965 (0.013)&   15.865    (0.015)&   15.725 (0.017) \\
2009/10/12 &   55116.3 &                &   16.798 (0.016)&   15.975 (0.015)&   15.864    (0.013)&   15.696 (0.015) \\
2009/10/18 &   55122.3 &                &                &   15.972 (0.011)&   15.855    (0.012)&   15.695 (0.014) \\
2009/10/24 &   55128.2 &                &   16.804 (0.014)&                &                   &                 \\
2009/10/28 &   55132.3 &                &   16.807 (0.014)&   15.976 (0.015)&   15.847    (0.012)&   15.695 (0.015) \\
2009/11/02 &   55137.2 &                &   16.811 (0.014)&   15.974 (0.013)&   15.841    (0.013)&   15.704 (0.017) \\
2009/11/05 &   55140.3 &                &   16.826 (0.015)&   15.979 (0.015)&   15.843    (0.017)&   15.696 (0.025) \\
2009/11/08 &   55143.2 &                &   16.825 (0.018)&                &                   &                 \\
2009/11/11 &   55146.2 &                &   16.840 (0.018)&                &                   &                 \\
2009/11/13 &   55148.2 &                &   16.865 (0.016)&                &                   &                 \\
2009/11/16 &   55151.2 &                &                &   16.006 (0.013)&   15.881    (0.019)&   15.740 (0.019) \\
2009/11/24 &   55159.2 &                &                &   16.051 (0.015)&   15.940    (0.016)&   15.761 (0.015) \\
2009/12/03 &   55168.1 &                &   17.140 (0.020)&                &                   &                 \\
2009/12/07 &   55172.2 &                &   17.259 (0.017)&   16.229 (0.013)&   16.146    (0.013)&   15.927 (0.019) \\
2009/12/09 &   55174.2 &                &   17.326 (0.019)&                &                   &                 \\
2009/12/10 &   55175.2 &                &   17.371 (0.026)&                &                   &                 \\
2009/12/11 &   55176.2 &                &                &                &   16.428    (0.032)&                 \\
2009/12/12 &   55177.2 &                &   17.474 (0.022)&                &                   &                 \\
2009/12/14 &   55179.2 &                &   17.603 (0.021)&                &                   &                 \\
2009/12/15 &   55180.2 &                &   17.689 (0.021)&                &                   &                 \\
2009/12/18 &   55183.1 &                &   17.985 (0.030)&                &                   &                 \\
2009/12/20 &   55185.1 &                &   18.198 (0.030)&                &                   &                 \\
2009/12/21 &   55185.3 &                &                &   16.981 (0.024)&   16.854    (0.028)&   16.556 (0.032) \\
2009/12/23 &   55188.1 &                &   18.398 (0.030)&                &                   &                 \\
2009/12/24 &   55189.1 &                &   18.395 (0.030)&                &                   &                 \\
2009/12/26 &   55191.1 &                &                &   17.184 (0.022)&   17.039    (0.028)&   16.715 (0.024) \\
2009/12/27 &   55192.1 &                &                &   17.222 (0.021)&   17.077    (0.028)&   16.735 (0.028) \\
2009/12/28 &   55193.2 &                &                &   17.316 (0.025)&   17.119    (0.030)&   16.768 (0.035) \\
2010/01/01 &   55197.1 &                &   18.419 (0.025)&                &                   &                 \\
2010/01/02 &   55198.1 &                &                &   17.361 (0.021)&   17.176    (0.020)&   16.878 (0.020) \\
2010/01/04 &   55200.1 &                &   18.494 (0.024)&                &                   &                 \\
2010/01/13 &   55209.0 &                &   18.498 (0.032)&                &                   &                 \\
2010/01/18 &   55214.1 &                &   18.576 (0.035)&                &                   &                 \\
2010/02/13 &   55240.1 &                &   18.742 (0.027)&                &                   &                 \\
\hline
\end{tabular}			  
\end{minipage}
\end{table*}

\begin{table*}
 \centering
 \begin{minipage}{84mm}
  \caption{$JH$ magnitudes of SN 2009ib.}
  \label{lc_nir}
  \begin{tabular}{@{}cccc@{}}
  \hline
  \hline
Date & MJD & $J$ & $H$  \\
  \hline
2009/08/12 & 55056.3 &  15.36 (0.22) & 15.08 (0.56) \\
2009/08/20 & 55063.3 &  15.15 (0.27) & 15.05 (0.67) \\
2009/08/25 & 55068.3 &  15.13 (0.18) & 14.85 (0.63) \\ 
2009/08/30 & 55073.4 &  15.06 (0.11) & 14.66 (0.24) \\
2009/09/10 & 55084.2 &  14.91 (0.15) & 14.57 (0.45) \\
2009/09/15 & 55089.3 &  14.92 (0.07)  &14.51 (0.56) \\ 
2009/09/21 & 55095.2 &  14.87 (0.16) & 14.48 (0.12) \\
2009/09/26 & 55100.2 &  14.84 (0.22) & 14.51 (0.64) \\
2009/10/12 & 55116.3 &  14.67 (0.68) & 14.45 (0.11) \\ 
2009/10/17 & 55121.3 &  14.77 (0.18) & 14.45 (0.26) \\
2009/10/23 & 55128.2 &  14.73 (0.05) & 14.43 (0.35) \\
2009/11/06 & 55141.2 &  14.76 (0.17) & 14.47 (0.37) \\
2009/11/11 & 55146.3 &  14.72 (0.12) & 14.50 (0.36) \\
\hline
\end{tabular}			  
\end{minipage}
\end{table*}

\end{document}